\documentclass[%
 reprint,
superscriptaddress,
nofootinbib,
 amsmath,amssymb,
 aps,
]{revtex4-2}

\usepackage{graphicx}
\usepackage{dcolumn}
\usepackage{bm}
\usepackage{float} 

\usepackage[dvipsnames]{xcolor}
\usepackage[normalem]{ulem}

\usepackage[caption=false]{subfig} 

\usepackage{appendix}

\begin{document}

\preprint{APS/123-QED}

\title{Optimal Transport for $e/\pi^0$ Particle Classification in LArTPC Neutrino Experiments}

\author{David Caratelli}
\email{dcaratelli@ucsb.edu}
 \affiliation{University of California, Santa Barbara}
 \author{Nathaniel Craig}
\email{ncraig@ucsb.edu}
 \affiliation{University of California, Santa Barbara}%
 \affiliation{Kavli Institute for Theoretical Physics, Santa Barbara, CA USA}
\author{Chuyue Fang}%
 \email{cfang@ucsb.edu}
\affiliation{University of California, Santa Barbara}
\author{Jessica N. Howard}%
 \email{jnhoward@kitp.ucsb.edu}
\affiliation{Kavli Institute for Theoretical Physics, Santa Barbara, CA USA}

\date{\today}

\begin{abstract}
The efficient classification of electromagnetic activity from $\pi^0$ and electrons remains an open problem in the reconstruction of neutrino interactions in Liquid Argon Time Projection Chamber (LArTPC) detectors. We address this problem using the mathematical framework of Optimal Transport (OT), which has been successfully employed for event classification in other HEP contexts and is ideally suited to the high-resolution calorimetry of LArTPCs. Using a publicly available simulated dataset from the MicroBooNE collaboration, we show that OT methods achieve state-of-the-art reconstruction performance in $e/\pi^0$ classification. The success of this first application indicates the broader promise of OT methods for LArTPC-based neutrino experiments.
\end{abstract}

\maketitle

\section{\label{sec:intro} Introduction}
The observation of neutrino oscillations~\cite{Huber_2022lpm} is currently one of the most compelling motivations for the existence of non-Standard Model physics, making a better understanding of neutrino properties and interactions a crucial goal of particle physics in the coming years. Ongoing and future experiments such as those which are part of the Short Baseline Neutrino (SBN) program~\cite{bib:SBN} or the Deep Underground Neutrino Experiment (DUNE)~\cite{bib:DUNETDR,bib:DUNEOSC} rely on Liquid Argon Time Projection Chamber (LArTPC) detectors for granular imaging and calorimetry of particle interactions over large volumes.
It is therefore important to explore and understand potential data analysis strategies that can meet and expand on the physics goals of future LArTPC experiments.

A main feature of LArTPCs is their significantly enhanced resolution capabilities compared to other neutrino detector strategies~\cite{bib:MiniBooNE,bib:NOvA,bib:MINOS}. 
This increased resolution is often not fully exploited by traditional analysis techniques. Many groups~\cite{Drielsma_2021jdv,bib:NuGraph,bib:DUNECNN,bib:LArTPCgenerative,bib:DL1,bib:DL2} have explored using Machine Learning (ML) to improve the data analysis pipeline. 
While powerful, ML can also present difficulties related to interpretability and uncertainty quantification.

Optimal Transport (OT) is a mathematical framework that furnishes a notion of the distance between two probability distributions in terms of the minimal ``work'' required to rearrange one into another. In recent years OT methods have proven powerful for image processing, which suggests their relevance to data analysis in high energy physics (HEP) experiments. Indeed, OT methods have been fruitfully applied to the analysis of Large Hadron Collider (LHC) 
data~\cite{hepmllivingreview, 
ATLAS:2025rbr,
Algren:2023spv,
Leigh:2024chm,
Manole:2022bmi,
Davis:2023lxq,
Cheng:2020dal,
Larkoski:2025pai,
Larkoski:2025idq,
Cai:2025fyw,
Craig:2024rlv,
Gambhir:2024ndc,
Gaertner:2023tzv,
Larkoski:2023qnv,
Ba:2023hix,
Gouskos:2022gjg,
Park:2022zov,
Cai:2021hnn,
Tsan:2021brw,
DiGuglielmo:2021ide,
Collins:2021pld,
Cai:2020vzx,
CrispimRomao:2020ejk,
Komiske:2019fks,
Komiske:2019jim,
Fraser:2021lxm,
Komiske:2020qhg,
Cesarotti:2024tdh,
ATLAS:2023mny,
Cesarotti:2020hwb,
Cesarotti:2020ngq}.
These applications have taken a variety of forms. For example, recently OT transport plans have been used to calibrate ATLAS flavour-tagging algorithms~\cite{ATLAS:2025rbr}, to decorrelate observables~\cite{Algren:2023spv}, generate templates~\cite{Leigh:2024chm}, and estimate backgrounds~\cite{Manole:2022bmi}. The majority of applications of OT to data analysis rely on calculating distances between distributions of interest. For example, OT distances have been used on Dalitz distributions to measure CP violation~\cite{Davis:2023lxq} or on the loss functions of trained VAEs to help identify anomalies~\cite{Cheng:2020dal}. However, the applications most relevant to the current work choose to represent a collider event or object (e.g. jet) as a discrete distribution over an underlying ground space~\cite{Larkoski:2025pai,
Larkoski:2025idq,
Cai:2025fyw,
Craig:2024rlv,
Gambhir:2024ndc,
Gaertner:2023tzv,
Larkoski:2023qnv,
Ba:2023hix,
Gouskos:2022gjg,
Park:2022zov,
Cai:2021hnn,
Tsan:2021brw,
DiGuglielmo:2021ide,
Collins:2021pld,
Cai:2020vzx,
CrispimRomao:2020ejk,
Komiske:2019fks,
Komiske:2019jim,
Fraser:2021lxm,
Komiske:2020qhg,
Cesarotti:2024tdh,
ATLAS:2023mny,
Cesarotti:2020hwb,
Cesarotti:2020ngq}.
Subsequently, OT distances are calculated between these event representations~\cite{Larkoski:2025pai,
Larkoski:2025idq,
Cai:2025fyw,
Craig:2024rlv,
Gambhir:2024ndc,
Gaertner:2023tzv,
Larkoski:2023qnv,
Ba:2023hix,
Gouskos:2022gjg,
Park:2022zov,
Cai:2021hnn,
Tsan:2021brw,
DiGuglielmo:2021ide,
Collins:2021pld,
Cai:2020vzx,
CrispimRomao:2020ejk,
Komiske:2019fks,
Komiske:2019jim,
Fraser:2021lxm,
Komiske:2020qhg} or between the event representation and a constructed reference~\cite{Fraser:2021lxm,
Komiske:2020qhg,
Cesarotti:2024tdh,
ATLAS:2023mny,
Cesarotti:2020hwb,
Cesarotti:2020ngq} to define a similarity metric. 
Note that for a certain choice of cost-function the latter can implicitly approximate the pairwise distances between events~\cite{Cai:2020vzx, Cai:2021hnn}.

In general, OT distances between events can be used for event classification and anomaly detection on their own or coupled to simple distance-based machine learning methods. The performance of these methods is often competitive with standard deep learning approaches, but more amenable to interpretation and uncertainty quantification. In this respect, OT provides a promising foundation for interpretable, physics-informed machine learning methods in HEP.

The success of OT methods applied to LHC datasets naturally suggests their relevance to other HEP calorimeters. In this work, we explore how OT can be used for data analysis in LArTPC experiments. As a test case, we explore the performance of OT on the task of $e/\pi^0$ classification. Neutral pion background rejection is particularly challenging for current algorithms, leading to significant backgrounds in analyses with electron~\cite{bib:uB_eLEE} and single-photon~\cite{bib:uB_ee,bib:uB_singlephoton} final-states.
We find that OT distances can be used to yield state-of-the art classification performance on the task of $e/\pi^0$ separation. This is particularly true when OT distances are combined with Support Vector Machines (SVMs), an interpretable ML method. More generally, we find that all OT-based methods explored in this work significantly outperform traditional reconstruction algorithms. 

This paper is organized as follows. Section \ref{sec:detector} describes the LArTPC technology and the physics that LArTPC detectors located in neutrino beams are exploring. Sec.~\ref{sec:detector_simulation} describes the simulation and low-level reconstruction that are used to produce Monte Carlo samples for analysis, and in particular discusses the ``MicroBooNE Open Dataset'' used by this work for accurate rendering of neutrino images in a LArTPC. Sec.~\ref{sec:the_problem} discusses the task of $e/\pi^0$ classification, which is the focus of this work. Section~\ref{sec:OT} reviews OT distances and motivates our particular choice (balanced 2-Wasserstein distances) in the context of $e/\pi^0$ classification. Section~\ref{sec:data_processing} overviews the data pre-processing, and Sec.~\ref{sec:methods} discusses the post-processing methods employed in this work. The results are presented in Sec.~\ref{sec:results}, followed by a brief conclusion which discusses directions for future work in Sec.~\ref{sec:conclusion}.

\section{\label{sec:detector} Liquid Argon Time Projection Chamber Neutrino Detectors}

\subsection{Physics Landscape with LArTPC Detectors}

Because of their imaging capabilities, LArTPC detectors have become the technology of choice for HEP experiments operating in accelerator-based neutrino beams. LArTPCs are able to capture images of particle interactions with millimeter position accuracy and percent-level calorimetric resolution~\cite{bib:protoDUNEdetector,bib:uBcalibration,bib:ICARUSdetector}. This, in turn, allows for excellent particle identification (PID) which is leveraged to perform precision measurement of neutrino oscillations~\cite{bib:uB_sterile} and searches for BSM particles~\cite{bib:uB_HNL_2020,bib:uB_HNL_2022,bib:uB_DarkTrident}. These detectors complement other technologies employed for accelerator-based neutrino experiments, including water-Cherenkov detectors such as the next-generation Hyper-Kamiokande~\cite{bib:hyperk} experiment.

Common final states in LArTPCs exposed to neutrino beams consists of electromagnetic showers produced by $\mathcal{O}$(0.01-1) GeV electrons and photons. Electron final-states are the key signature of $\nu_{\mu} \rightarrow \nu_e$ oscillations being studied with both the DUNE and SBN programs. 
Another important component of this physics program are BSM signatures in the sub-GeV regime which are generally tied to ``Dark Sector'' models~\cite{bib:DarkSector1,bib:DarkSector2}. These models predict the production of new particles along the neutrino beamline. Because of the energy regime, these new particles often lead to final-state observables consisting of $e^+e^-$ or photon electromagnetic (EM) showers.
At the same time, photon EM showers are abundantly produced in standard model (SM) neutrino interactions, primarily through resonant pion production where a $\nu$-Ar interaction excites a $\Delta(1232)$ or heavier baryon resonance~\cite{bib:RES1,bib:RES2} which can decay to a final-state $\pi^0$ producing two $\gamma$ showers in the detector. Roughly 10\% of neutrino interactions lead to a $\pi^0$ final-state, far outnumbering the predicted signatures of neutrino oscillations or BSM signatures.
Because of this, the physics program that these detectors hope to achieve relies critically on $e$-$\gamma$ and $e$-$\pi^0$ separation.

\subsection{Introduction to LArTPC detectors}

\begin{figure}[h]
    \begin{center}
    \includegraphics[width=0.9\columnwidth]{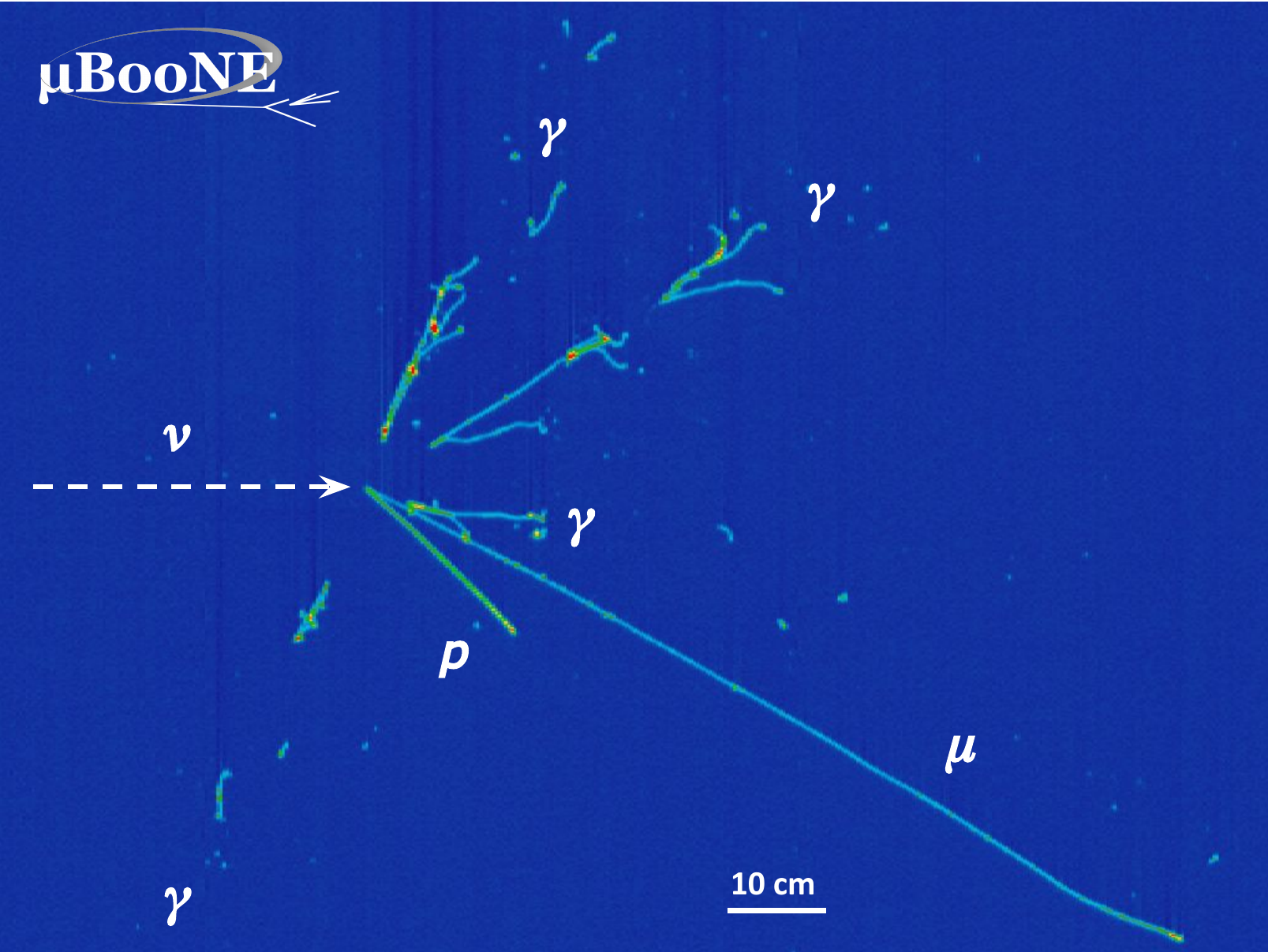}
    \end{center}
\caption{Example event display of a neutrino interaction collected with one wire plane of the MicroBooNE LArTPC detector. Four photon EM showers, likely from the decay of two $\pi^0$ mesons, can be seen branching out from the interaction vertex. The x-axis represents the wire position while the y-axis indicates the time taken for the electrons to drift to the wire plane. The color scale in the event display shows the measured charge deposit.}
\label{fig:evd}
\end{figure}

Charged particles produced in neutrino interactions propagate through the LArTPC, ionizing and exciting the argon atoms that make up the active mass of the detector. An external electric field causes ionization electrons to drift uniformly from the bulk of the detector to the positive potential wall (anode) of the TPC. Wires~\cite{bib:uBdetector,bib:ICARUSdetector,bib:protoDUNEdetector},  strips~\cite{bib:dune-vd}, or pixel-based~\cite{bib:2x2operations,bib:dune-module0} charge readout sensors measure the induced current caused by the propagating electron clouds. Charge reaches these sensors at different times, depending on the exact location where the energy was deposited. In the case of wire- or strip-based detectors, multiple overlapping planes allow for the formation of different 2D views of the charge deposition, which are in turn used to reconstruct a 3D image of the interaction~\cite{bib:uBpandora,bib:uBWC1}. Pixel-based detectors inherently provide 3D reconstructed charge deposits.
In the GeV energy-scale range, different particle species give rise to different characteristic topologies. Particles such as protons, charged pions, or muons lose energy primarily through ionization, producing linear ``track-like'' signatures in the detector. Electrons and photons, on the other hand, lose energy via ionization and radiation in roughly equal amounts, leading to the formation of electromagnetic (EM) showers which appear as ``fuzzy'' branches and $e^+e^-$ pair cascades. Below one GeV, these EM showers are often highly segmented due to the stochastic nature of radiative energy loss. A more in-depth description of EM signatures in LArTPCs can be found in Refs.~\cite{bib:uBMichel,bib:uBpi0}. Figure~\ref{fig:evd} shows an example neutrino interaction image collected by the MicroBooNE LArTPC, where different final-state particles can be seen producing topologically distinct signatures.

\section{\label{sec:detector_simulation} Detector Simulation and Low-Level Reconstruction}

In this section we describe general features of the simulation for accelerator-based LArTPC detectors as well as the details of the dataset used specifically in this work~\cite{bib:uBopendataincl, bib:uBopendatanue}. We also describe the 3D reconstruction applied to produce the inputs used in our analysis.

\subsection{MicroBooNE Open Dataset}
This work uses Monte Carlo simulations of electron and $\pi^0$ interactions obtained from the MicroBooNE Open Dataset~\cite{bib:uBopendataincl,bib:uBopendatanue}. These samples were released by the MicroBooNE collaboration with the purpose of encouraging broader community participation and expanding the development of reconstruction and analysis methods for LArTPC detectors. The MicroBooNE Open Dataset consists of a sample of $10^6$ simulated $\nu_{\mu}$ and $10^5$ $\nu_e$ interactions. In addition to providing a high-statistics and easily-accessible dataset, these samples are the same used by the MicroBooNE collaboration in physics analyses, and incorporate noise and other detector effects that help provide a realistic modeling of the detector response.
The results of this work therefore can be interpreted as a realistic assessment of the performance to be expected when analyzing real data in a LArTPC experiment, enhancing the robustness and value of the work presented.

\subsection{Detector Simulation and Calibrations}

The MicroBooNE simulation comprises three main components: 
\begin{itemize}
    \item The Booster Neutrino Beamline flux~\cite{bib:bnb}, a $\sim$1 GeV neutrino energy beamline comprised of 95\% $\nu_{\mu}$, 4\% $\bar{\nu}_{\mu}$, and $< 1\%$ $\nu_e$/$\bar{\nu}_e$. 
    \item The simulation of neutrino interactions with the detector's argon target. This is carried out with the GENIE event generator~\cite{bib:GENIE} using a suite of neutrino interaction models and tailored tunes described in Ref.~\cite{bib:ubtune}.
    \item The simulation of the detector response, which is carried out through custom code in the LArSoft framework~\cite{bib:larsoft}. This stage accounts for ionization charge propagation in the detector and signal formation on TPC wires.
\end{itemize}

\begin{figure}[h]
\includegraphics[width=0.8\columnwidth]{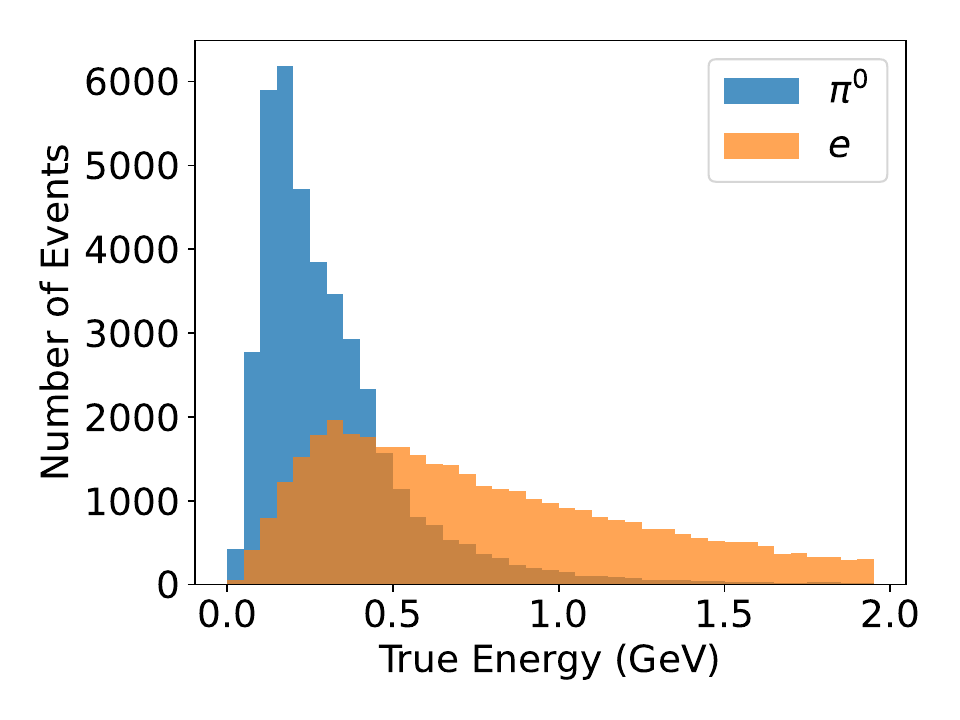}
\caption{\label{fig:epi0energy} True energy distribution of $e$ and $\pi^0$ particles in the MicroBooNE open datasets used for this work.}
\end{figure}

The first two components of the simulation primarily influence the kinematics of particles produced in the detector from $\nu$-Ar interactions. For the purpose of the particle classification task being performed in this work, we emphasize the characteristics of electron and $\pi^0$ events. Electron neutrinos, interacting through charged-current interactions, produce electrons with an energy spectrum which spans the 0.1-2 GeV range, peaking at approximately 0.5 GeV. Neutral pions, on the other hand, are generally produced with low momentum, peaking at 200 MeV. Figure~\ref{fig:epi0energy} shows the true energy distribution of electrons and $\pi^0$ particles produced in neutrino interactions simulated in the MicroBooNE open datasets used for this work.

The last component of the simulation is responsible for incorporating all realistic  effects that impact the detector.
Effects that alter the amount of charge collected, or its propagation through the detector such as ion recombination, space-charge, and diffusion are simulated based on data-driven measurements performed in-situ or with dedicated experiments~\cite{bib:uBSCE}. A set of calibration steps are finally applied to MicroBooNE data and simulation alike to account  and correct for detector distortions. These are described in Ref.~\cite{bib:uBcalibration}.

\subsection{Low-Level Reconstruction}

Signals recorded on TPC wires first undergo low-level reconstruction which is responsible for noise filtering~\cite{bib:ubnoise} and signal processing~\cite{bib:ubsignal1,bib:ubsignal2}. These lead to the formation of reconstructed objects called ``hits'', which represent individual pixels in images such as those of Fig.~\ref{fig:evd}, and convey the position (wire number, on the horizontal axis), drift time to reach the wire (vertical axis), and charge measured on the wire (color-scale). Hits are reconstructed independently on each plane and form the basis for most LArTPC reconstruction workflows. 

\subsection{Current Status of LArTPC Pattern Recognition}

After the identification of hits on each plane, different reconstruction workflows then diverge in their approach. Two commonly used reconstruction paradigms are the \texttt{Pandora} multi-algorithm workflow~\cite{bib:uBpandora,bib:protoDUNEpandora} and the \texttt{Wire-Cell} toolkit~\cite{bib:uBWC1,bib:uBWC2}. Pandora uses 2D hit inputs to perform tracking and 3D particle imaging. Wire-Cell instead uses 2D inputs to reconstruct directly 3D ionization charge patterns through a tomographic approach, after which higher level pattern-recognition techniques are employed for particle track reconstruction. These reconstruction frameworks have been developed over a number of years with data from the MicroBooNE~\cite{bib:uBdetector} and protoDUNE~\cite{bib:protoDUNEdetector} detectors, and are now widely deployed across active LArTPC detectors including SBND and ICARUS. Finally, current ML approaches often directly use the 2D images made up by wire signals or reconstructed hits to perform high-level particle ID and kinematics reconstruction. These ML tools often rely on convolutional neural networks (CNNs) and graphical neural networks (GNN) to perform specific image-classification tasks that aim to classify features of LArTPC images at either particle-level~\cite{bib:uBCNN1,bib:uBCNN2,bib:DUNECNN1,bib:DUNECNN2} or pixel by pixel~\cite{bib:uBSS1,bib:uBSS2}
or perform high-level reconstruction tasks such as energy reconstruction~\cite{bib:uBDLereco}.

\section{\label{sec:the_problem} $e/\pi^0$ Particle Classification}

As described in Sec.~\ref{sec:detector}, electrons and photons are key final-state observables for neutrino oscillations and BSM signatures in LArTPC experiments, but they are flooded by large backgrounds from $\pi^0$ decays to di-photons.
 One of the hallmarks of LArTPC detectors in neutrino experiments is their ability to distinguish electron from photon showers via two key features: (1) calorimetric d$E$/d$x$ separation of $e$ vs. $\gamma$ due to the rate of energy deposition at the start of the EM shower, and (2) the presence of a ``gap'' from the interaction vertex to the $\gamma$ EM shower starting point due to the $\sim$20 cm conversion distance in argon. These two features allow for powerful background rejection~\cite{bib:ArgoNeuT_dedx,bib:uB_dedx,bib:uBeLEE2025}. Yet, because the vast majority of $\gamma$ backgrounds originate from $\pi^0$s, there is another, more fundamental distinguishing feature: $\pi^0$ events lead to two EM showers, one from each photon. In principle, $\pi^0$ backgrounds could be effectively mitigated by correctly identifying both photon showers. Successfully identifying both EM showers nonetheless poses several challenges. The $\pi^0 \rightarrow \gamma\gamma$ decay leads to two photons which may be collinear, and therefore hard to differentiate. In other cases, a very asymmetric decay leads to one of the photons having little energy. This is very common, in part because $\pi^0$s produced in neutrino interactions typically have a low momentum, causing the sub-leading photon in the decay to often have less than 100 MeV of kinetic energy. Finally, sometimes photons from the decay escape the detector before depositing energy, which precludes identifying the second EM shower.
Importantly, identifying \emph{both} photons from a $\pi^0$ decay remains a significant challenge in LArTPC detectors, and is the primary limitation to further reducing $\pi^0$ backgrounds which currently limit the sensitivity of many LArTPC physics measurements~\cite{bib:uB_ee,bib:uB_singlephoton}. In this section we will explore the current status of $e$/$\pi^0$ separation in LArTPC detectors, and motivate how OT can improve this.

\begin{figure}[ht]
    \subfloat[\label{fig:EVDPi0}]{%
        \includegraphics[width=0.85\columnwidth]{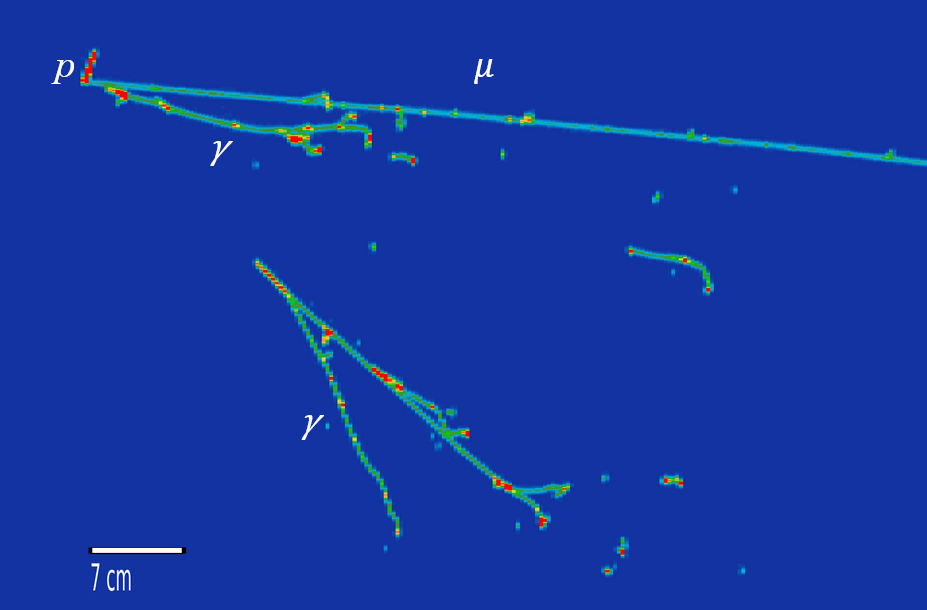}
    }\hspace*{\fill}\\
    \subfloat[\label{fig:EVDElectron}]{%
        \includegraphics[width=0.85\columnwidth]{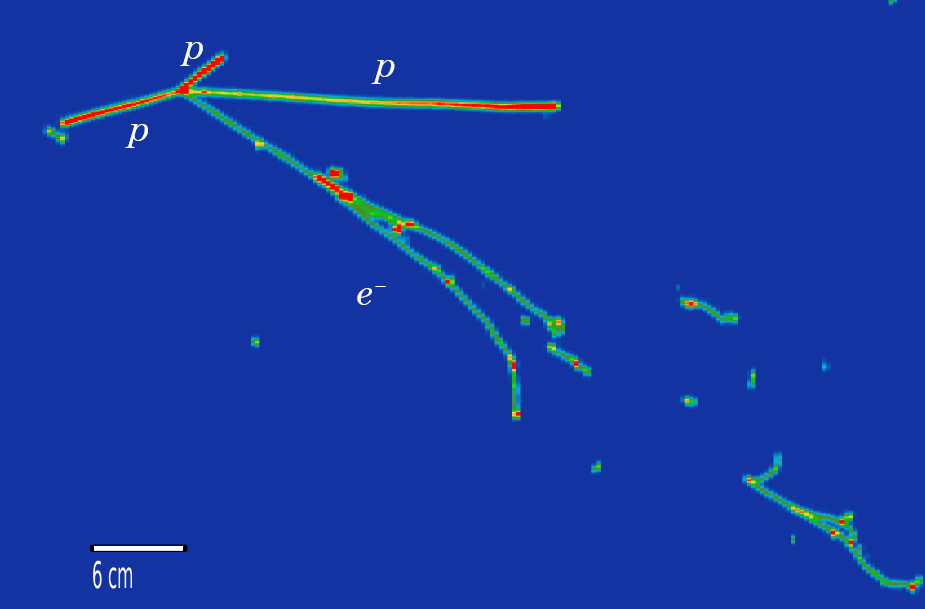}
    }%
    \caption{Example event displays for $\pi^0$ (a) and $\mathrm{e}^-$ (b).}
    \label{fig:EVD-Events}
\end{figure}

\subsection{Current Status of LArTPC Reconstruction for $e/\pi^0$}

Many physics results produced with LArTPC neutrino experiments leverage the Pandora, Wire-Cell, or ML tools for electron-photon separation and $\pi^0$ rejection~\cite{bib:PeLEE,bib:uBNCCOH,bib:uBNCDEL,bib:uB_singlephoton,bib:uB_eLEE,bib:ArgoNeuT_dedx}. While these have demonstrated good performance, powerful $\pi^0$ rejection has come at the cost of a loss of efficiency that can negatively impact an analysis' sensitivity. As an example, $\nu_e$ efficiencies for recent analyses in MicroBooNE hover around $10-40$\%, with room for improvement. Furthermore, specific topologies such as isolated single-electrons with no hadronic activity~\cite{bib:PeLEE}, or more recent BSM searches for single-photons~\cite{bib:uBNCCOH,bib:uBNCDEL,bib:uB_singlephoton} are still heavily dominated by $\pi^0$ backgrounds to the point that this is the primary limitation to the analysis. Novel ways of rejecting $\pi^0$ backgrounds can therefore help expand the physics reach of analyses targeting single-electron or single-photon final-states. This is particularly true for methods that can overcome the reconstruction inefficiencies that often lead to not reconstructing one of the two photons from a $\pi^0$ decay.

Some of the ML tools developed aim to perform particle identification directly through a CNN, without the need to actually reconstruct the individual particles produced in an interaction (e.g. Ref.~\cite{bib:uBCNN2}). Such tools have been used to distinguish electrons from photons and other particles, but have not tackled the $e$/$\pi^0$ classification task. This paper explores OT as a new tool to tackle single-electron vs. $\pi^0$ classification through an image-based interpretation of the data, without the need to identify and reconstruct the two photons individually. The approach therefore is novel in that it follows a philosophy similar to that of many ML applications, which ``bypass'' several low-level reconstruction steps which may introduce significant inefficiencies, but does so introducing a new methodology based on OT distances, and tackling a classification problem that remains a key bottleneck in LArTPC physics analyses.
Given the current lack of a ML baseline for the $e$/$\pi^0$ classification task, we instead compare our OT-based results to reconstruction-based strategies which are currently used in analyses~\cite{bib:uB_dedx,bib:uBeLEE2025,bib:uBnueXSEC}.

\section{\label{sec:OT} Optimal Transport Distances for LArTPC Detectors}
The granular, calorimeter nature of LArTPC detectors makes image analysis tools well-suited for classification tasks. In turn, distinguishing different particle species by their visual topology in the detector is a task well suited for OT, which can quantify the ``distance'' between individual interaction images, leveraging features intrinsic to the physics of particle energy loss in a TPC. Importantly, OT can carry out this classification without the need to perform low-level reconstruction tasks which often suffer from significant inefficiencies.

\subsection{Review of Discrete OT Distances}
OT distances yield a mathematically rigorous notion of distance between probability distributions. Let $\mathcal{E}=\sum_{i=1}^{n} p_i \delta_{x_i}$ and $\mathcal{E'}=\sum_{j=1}^{n'} p'_j \delta_{x'_j}$ be two discrete probability distributions (which we will refer to as ``events'') defined over the same underlying $d$-dimensional metric space.\footnote{Note that $n$ need not be equal to $n'$. However, for balanced OT distances the total probability mass must be the same in each event i.e. $\sum_{i=1}^{n} p_i = \sum_{j=1}^{n'} p'_j$ } In other words, the event, $\mathcal{E}$, is a collection of $n$ points in $d$-dimensional space each weighted with probability $p_i$. 
OT distances elevate the pairwise distances between the points in $\mathcal{E}$ and $\mathcal{E'}$ ($| x_i - x'_j |$ for all possible combinations of $i,j$) to a distance between the events $\mathcal{E}, \mathcal{E'}$.
Intuitively, this is done by calculating the ``work'' to transform $\mathcal{E} \rightarrow \mathcal{E'}$ by moving the individual probability point masses. 
Moving a single probability point mass with weight $p$ from $x$ to $y$ costs an amount proportional to the distance traveled, $x-y$, and the amount of probability mass, $p$, moved.  
The total cost (``work'') of moving all points in $\mathcal{E} \rightarrow \mathcal{E'}$ is the sum of these individual costs. 
The optimal transport distance is the minimum cost incurred over all possible transport plans.

In this work we consider a particular kind of discrete, balanced OT distance, the 2-Wasserstein distance~\cite{Cai:2020vzx}, which is defined as follows,
\begin{equation}
W_2(\mathcal{E},\mathcal{E'}) = \left[ \min_{\gamma \in \Gamma(\mathcal{E},\mathcal{E'})} \sum_{i,j} \gamma_{i,j} |x_i - x'_j|^2 \right]^2
\end{equation}
where $\Gamma(\mathcal{E},\mathcal{E'})$ is the set of all possible ways to transform $\mathcal{E} \rightarrow \mathcal{E'}$, so-called ``transport plans'', and $\gamma$ denotes a particular transport plan. In statistical terms, $\Gamma(\mathcal{E},\mathcal{E'})$ is the set of all possible joint distributions with marginals $\mathcal{E},\mathcal{E'}$. The $\gamma^*$ which minimizes the above expression is unique and is known as the optimal transport plan. 

In this work we focus on calculating discrete OT distances between events.\footnote{We note that there has been interesting work on semi-discrete~\cite{Cesarotti:2024tdh, ATLAS:2023mny, Cesarotti:2020hwb, Cesarotti:2020ngq} alternatives in which events are represented as a discrete distribution but distances are calculated to a continuous manifold. The choice of manifold can help accentuate features in the events. However, in practice, this continuous manifold must be approximated by a discrete distribution.} This is a natural choice since the energy deposits in an event can naturally be expressed as a discrete probability distribution over the detector volume. Furthermore, balanced 2-Wasserstein distances are one type of a broad array of discrete OT distances. In the next section we motivate this choice in the context of $e/\pi^0$ classification.

\subsection{$e/\pi^0$ Classification with OT}
In order to apply OT to the task of  $e/\pi^0$ classification, we first note that we can view the energy deposits as discrete probability distributions in 3D space.  The amount of energy deposited (intensity) at a given spacepoint, up to a normalization factor, becomes the amount of probability mass.
Since the shapes of these events are visually indicative of their identity, it's natural to wonder whether OT distances can help in this task. 

There are many varieties of OT distances that one might think to apply. We have chosen the 2-Wasserstein distance for its nice mathematical properties, such as the uniqueness of the optimal transport plan and the ability to linearize the calculation~\cite{Cai:2020vzx}. 
2-Wasserstein distance is also an example of a \emph{balanced} OT strategy, meaning that the events must be normalized to have the same total intensity, which allows OT to focus on topological differences between events.

Electrons produced in $\nu_e$ interactions tend to have higher energies compared to the neutral pions (see Fig.~\ref{fig:epi0energy}), making the total amount of charge deposited by a shower a potentially powerful classification metric to separate signal $\nu_e$ from background $\pi^0$ interactions. 
While an unbalanced OT strategy~\cite{Cai:2021hnn} would be able to leverage this feature directly, for this work we aim to focus on the topological discrimination of $e$ and $\pi^0$ events due to the spatial distribution of the energy loss, and not the total energy deposited. 
This could make the classification task more challenging, but allows us to focus on evaluating OT's performance on tasks where existing methods are lacking. 
To this end, since balanced OT may still have indirect access to information on the total energy due to its correlation with the number of intensity points, we separate events into nine energy bins. Particle classification is carried out comparing events within each energy bin, avoiding classification dependencies on the energy of the particles.\\

We note that we also explored a different OT variant, Gromov-Wasserstein distances~\cite{Mémoli_2011}, as it is inherently insensitive to isometries (i.e. rotations and translations) of points in the ground space. 
Therefore, this strategy could serve as an alternative to 3D alignment (discussed in the next section). In a limited run on a single energy bin, Gromov-Wasserstein was shown to have similar performance to 3D alignment but did not see the same boost from post-processing the OT distances with interpretable machine learning strategies. Due to this lower performance as well as the high computational cost of traditional Gromov-Wasserstein algorithms, we opted to use 3D alignment instead. Results of the limited run are shown and discussed further in Appendix~\ref{app:GWresults}.\\

\section{\label{sec:data_processing} Data Pre-processing}

Previous work~\cite{Komiske:2019fks} has shown that pre-processing data (e.g. through rotations) to minimize distributional variations corresponding to non-physical differences is crucial for optimal transport.
For example, a calculation of the optimal transport distance between a given event and that same event rotated by some angle will generally yield a non-zero result, even though they are physically the same event.
Pre-processing the data to account for such physical symmetries (e.g. by aligning all tracks along the same direction) helps emphasize physically-meaningful distributional differences. 
This section describes the pre-processing steps implemented for this work with the aim of boosting the performance of OT-based methods. This includes how the 3D reconstructions of the events are obtained and aligned. Additionally, inspired by the success of jet images at the LHC~\cite{Komiske:2019fks}, we also consider subsequent 2D projections of these 3D events along the principal axes. The removal of small, isolated energy deposits (see Appendix~\ref{app:AnomalyFiltering}) was also attempted, but was found to have no significant improvement.

\subsection{Event Filtering}
Events with missing or incomplete showers will contaminate the sample and have a negative effect on separation using OT distances. For instance, a $\pi^0$ event with a missing second shower (for example when one of the photons escapes the detector) would appear to look like a single-shower event. Thus, having these $\pi^0$ events in the sample would be misleading for evaluating OT's performance. For this reason, we filter out events with showers (or large parts of the showers) which lie outside of the detector volume. This is done by removing events with more than 10 charge deposits falling within a 5 cm range from detector boundary.

\begin{figure*}[ht]
    \subfloat[\label{fig:3DPi0}]{%
        \includegraphics[width=1.1\columnwidth]{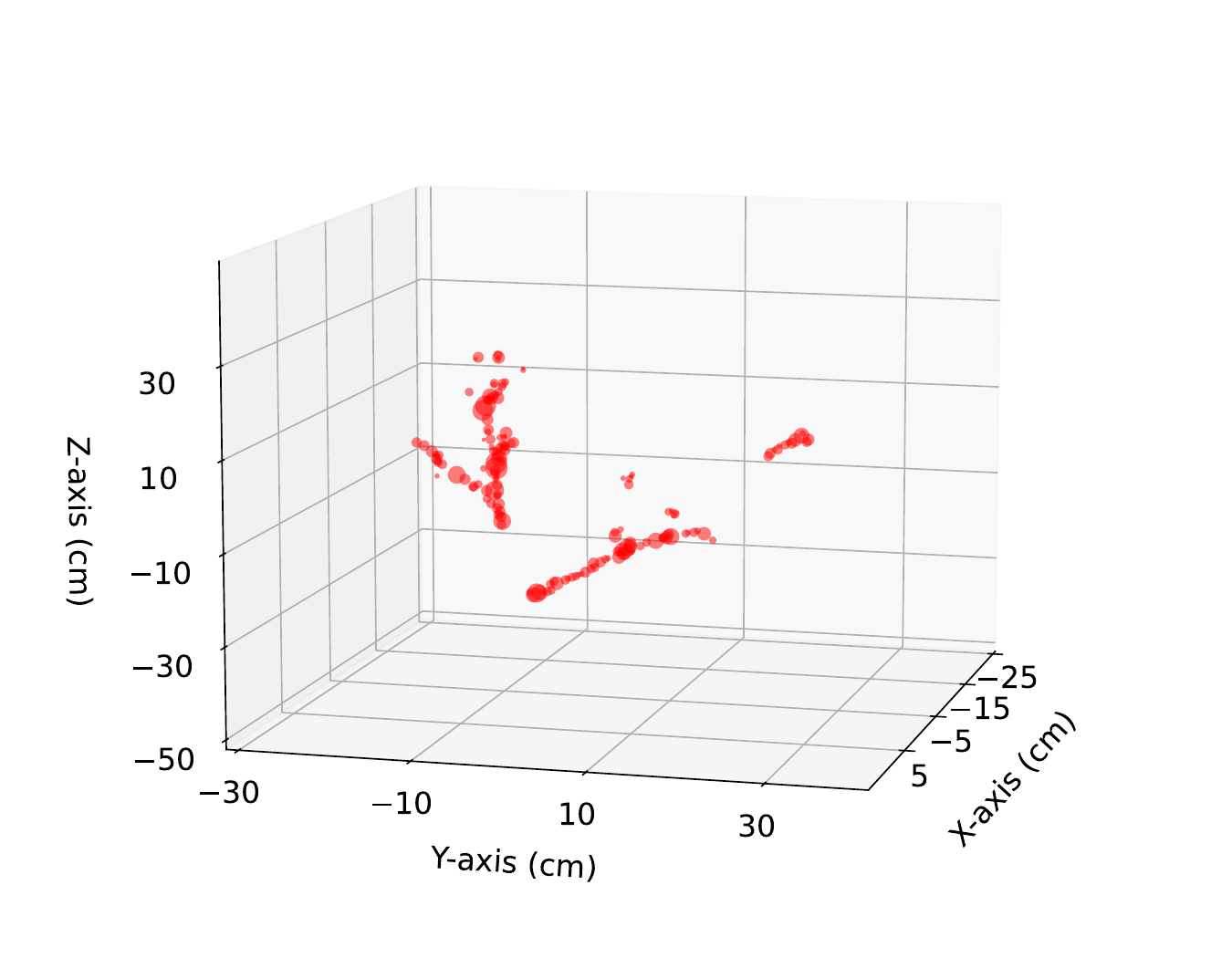}
    }\hspace*{\fill}
    \subfloat[\label{fig:3DElectron}]{%
        \includegraphics[width=1.1\columnwidth]{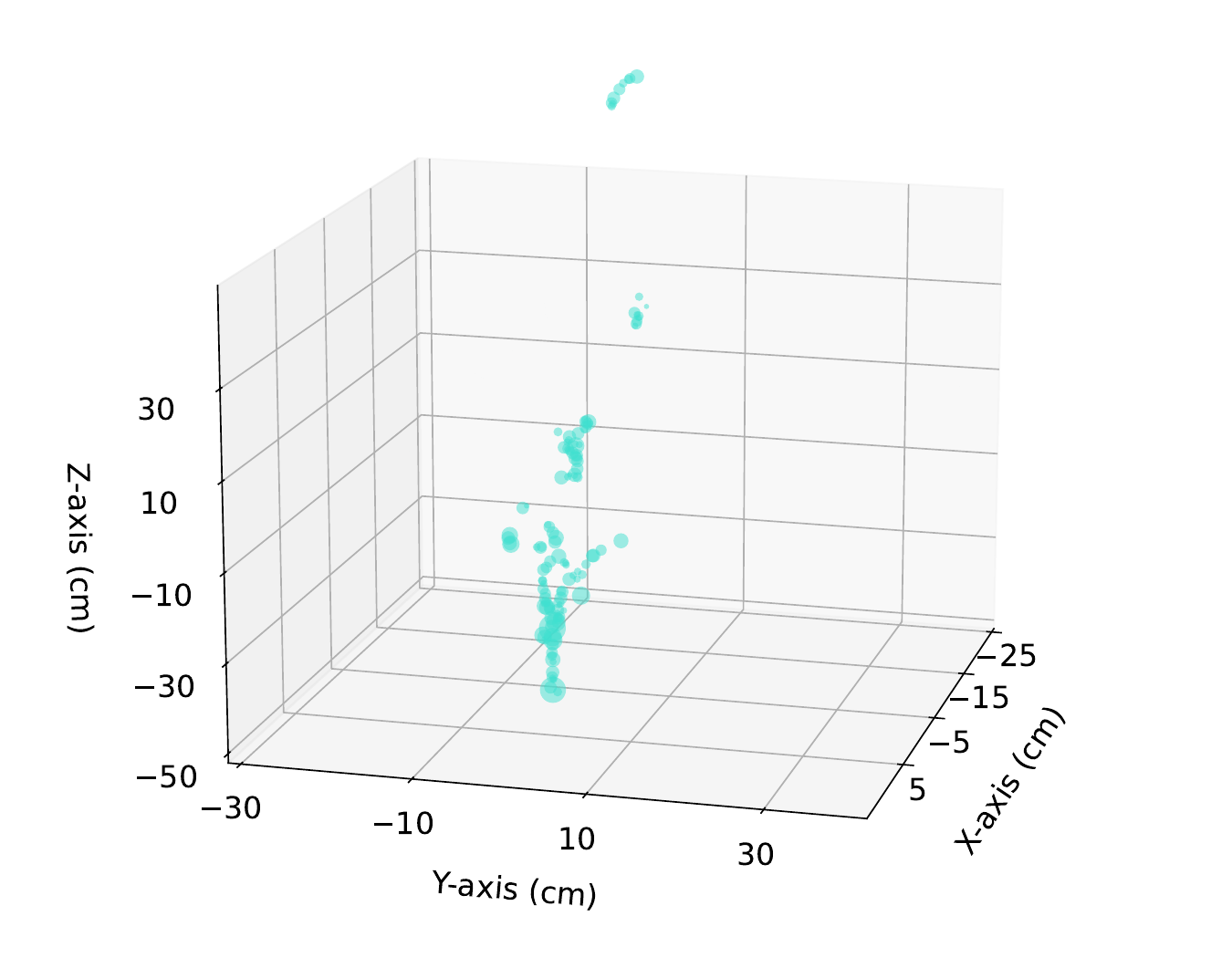}
    }%

    \caption{Events from Fig.~\ref{fig:EVD-Events} after 3D reconstruction and alignment. The size of the dots represent the relative amount of energy deposited associated with each reconstructed 3D spacepoint. (a) depicts a $\pi^0$ event and shows how the principal axis alignment based on the largest cluster in the event aligns the image with one of the two photon showers, keeping the second shower at an angle with respect to the $\hat{z}$ direction. (b) depicts an electron shower.}
    \label{fig:3D-Events}
\end{figure*}

\subsection{3D Reconstruction}
The three wire planes in MicroBooNE's LArTPC allow for three-dimensional (3D) reconstructions of the events. Since this work focuses on the supervised classification of $e$ from $\pi^0$ shower signatures, only hits associated with true electron or photon energy deposits are used for 3D reconstruction via the \texttt{SpacePointSolver} algorithm~\cite{abi2020dune}. The output of such reconstruction are spacepoints with position vectors in 3D space, each associated with two or three hits from different wire planes. Examples of this are shown in Figure \ref{fig:3D-Events}. The intensities of these spacepoints are determined using the charge deposits of their corresponding hits in the collection plane. 
These 3D reconstructions are the starting point for our classification analysis with OT. 
While for this work we rely on truth-labeling of EM showers to focus the evaluation of classification performance on the features that are most relevant to $e$ vs. $\pi^0$ classification, existing reconstruction methods would eventually be used to accomplish the task of isolating energy deposits associated with EM showers. ML methods which employ neural networks for hit-level semantic labeling can be integrated with OT for a wholistic reconstruction. Several such tools showing great promise have become available, including NuGraph~\cite{bib:NuGraph}, as well as other examples developed within the DUNE~\cite{bib:DUNECNN} and MicroBooNE~\cite{bib:uBSS1,bib:uBSS2} collaborations.

\subsubsection{Down Sampling}
Since the computational efficiency of OT depends on the number of spacepoints in an event, spacepoints within 1 cm distance of each other are merged into a single spacepoint with charge equal to the sum of the merged spacepoints. This down sampling is carried out to reduce the computation time. A 1 cm merging distance is small enough to still preserve the important features that distinguish EM showers. This reduces the number of spacepoints by a factor of $\sim$3, which results in an order of magnitude improvement in computational efficiency. See Sec.~\ref{sec:computationEfficiency} for more details on the computational cost scaling of OT distances.

\subsubsection{Spatial Alignment}
\label{sec:PCA}
Particles produced in neutrino interactions will propagate in different directions in the detector's coordinates. While a particle's orientation in the detector may carry valuable information on the kinematics of the interaction, it is not a feature that is relevant for particle classification, and may, in fact, complicate the particle classification task we wish to achieve. 
As previously described, pre-processing the data to account for these differences can enhance the ability for OT distances to pick up on the more physically relevant distributional differences.
Therefore, we first align the 3D distributions consistently by defining a common principal axis for each 3D reconstructed event.

To calculate the principal axis we use Weighted Principal Component Analysis (WPCA)~\cite{Delchambre_2014}. 
However, a subtlety arises when evaluating WPCA in the presence of the two showers in $\pi^0$ events:
for some events, using WPCA directly results in first principal components which are not representative of any of the shower axes.
To circumvent this issue we first cluster spacepoints together if they are within 2.5 cm distance of each other and compute WPCA on the cluster with the largest number of spacepoints.
The first principal component is then used to define the principal axis of the event.

Next, all events are rotated to align their principal axes, hence eliminating dependencies on the direction of the particle in the LArTPC. 
Similarly, a second rotation is carried out to align the second principal axes, in order to align the planar distribution of the events.
Events are then moved translationally such that the centers of mass are at the origin.
In Fig.~\ref{fig:3D-Events}, we show the electron and $\pi^0$ events from in Fig.~\ref{fig:EVD-Events} after 3D reconstruction and alignment.

\begin{figure*}[ht]
    \subfloat[\label{fig:3DPE-Transport}]{%
        \includegraphics[width=1.1\columnwidth]{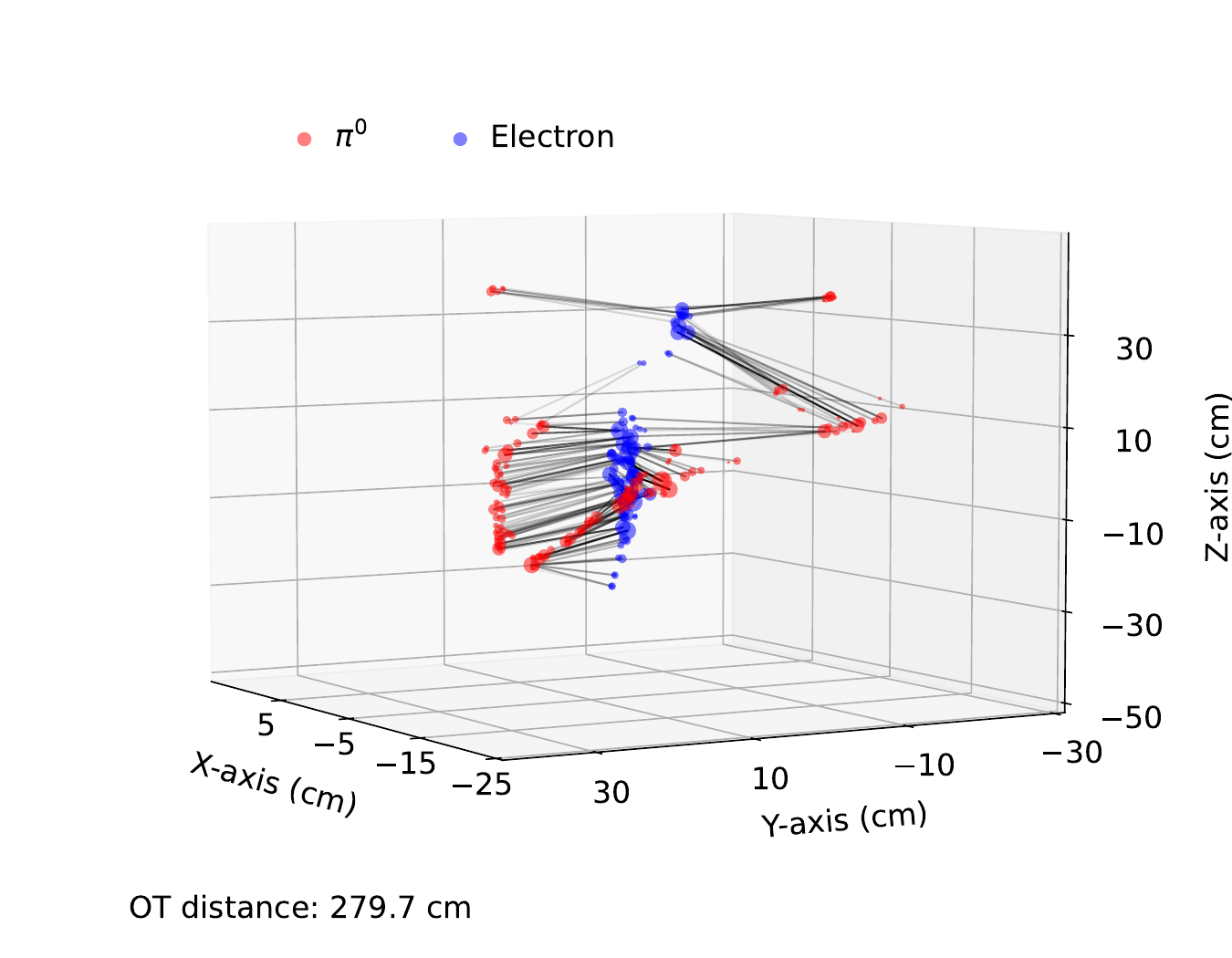}
    }\hspace*{\fill}
    \subfloat[\label{fig:3DEE-Transport}]{%
        \includegraphics[width=1.1\columnwidth]{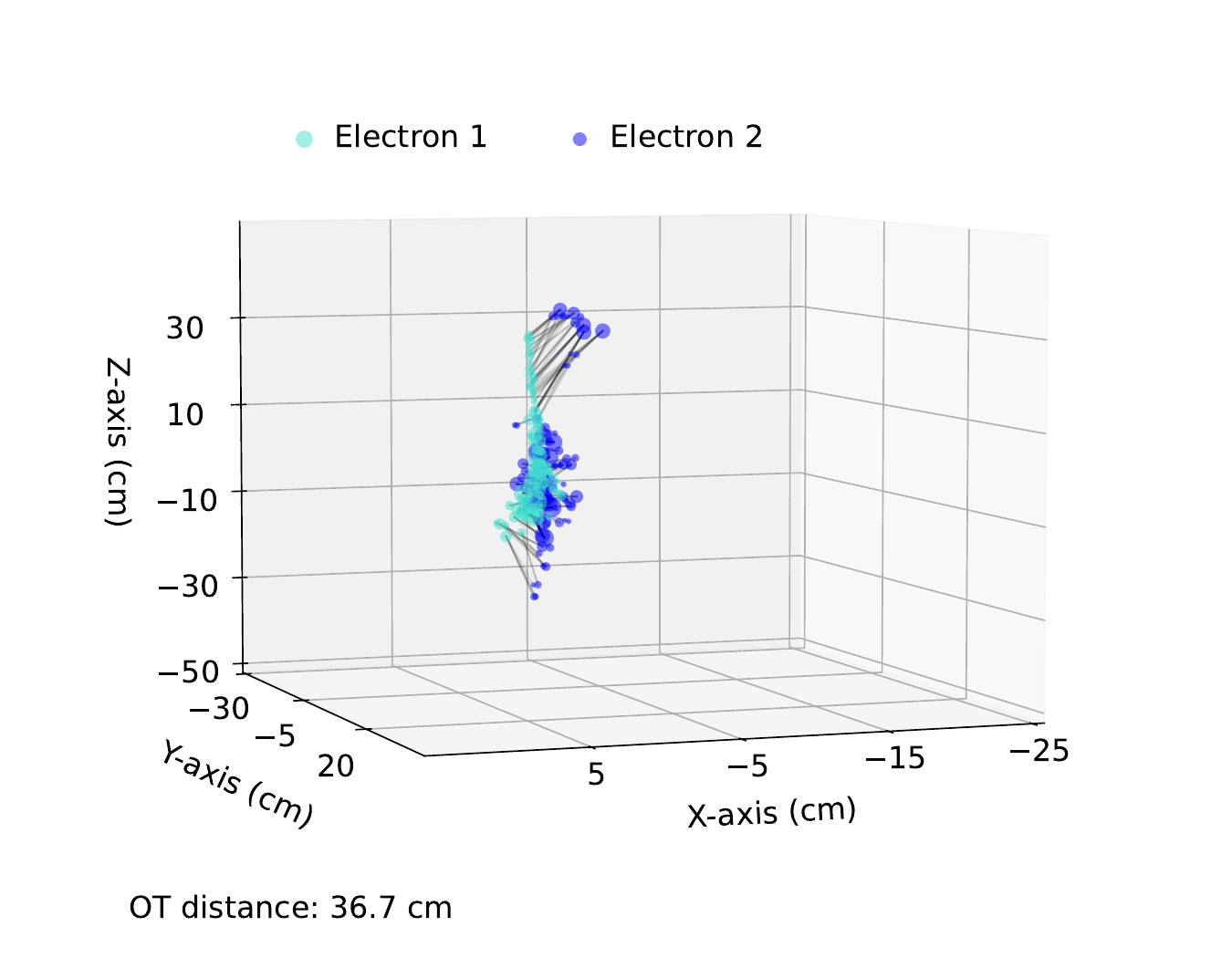}
    }%
    \caption{Example transport plans for optimal transport distances computed between two 3D event reconstructions. (a) transport plan between $\pi^0$ event (red) vs. ${e}^-$ event (blue); (b) transport plan between the same ${e}^-$ event (blue) and another $\mathrm{e}^-$ event (turquoise).}
    \label{fig:3D-Transport-Plans}
\end{figure*}

After these pre-processing steps, the OT distances can be computed. Two examples are shown in Fig.~\ref{fig:3D-Transport-Plans}, where a $\pi^0$ and an electron event are each transported into the same electron event shown in blue. The black lines show the visualized transport plans~\cite{Komiske:2019fks}, and the corresponding OT distances are computed. We can see that the $\pi^0$ in Fig.~\ref{fig:3DPE-Transport} takes a larger OT distance (279.7 cm) to transport all of its charge deposits into the electron event compared to the comparison between the two electrons of Fig.~\ref{fig:3DEE-Transport} (distance of 36.7 cm).

\begin{figure*}[hbt]
    \subfloat[\label{fig:2DPE-Transport}]{%
        \includegraphics[width=1.05\columnwidth]{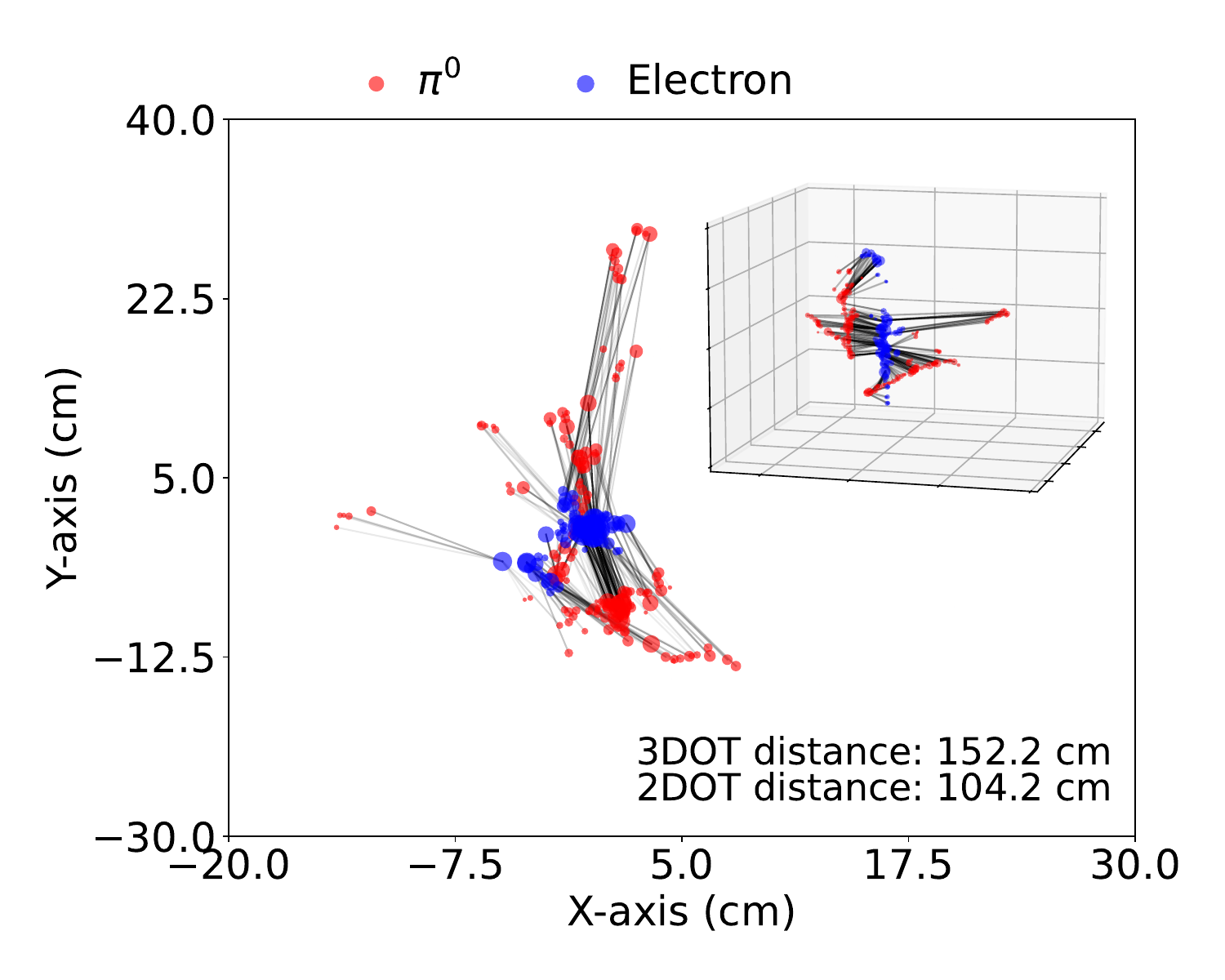}
    }\hspace*{\fill}
    \subfloat[\label{fig:2DEE-Transport}]{%
        \includegraphics[width=1.05\columnwidth]{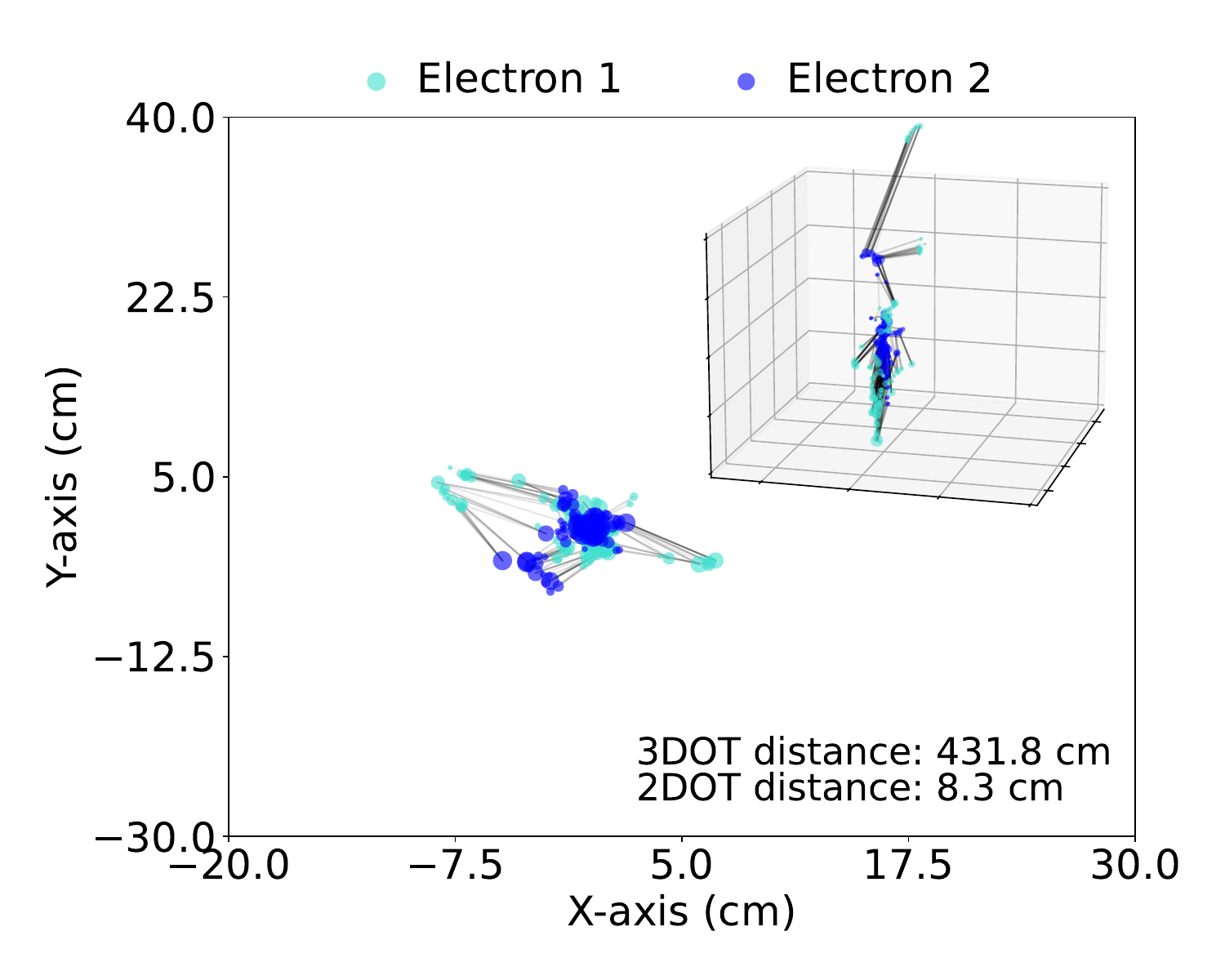}
    }%
    \caption{Transport plans for optimal transport computed between 2D projections (main figure) and 3D reconstructions (inset) of the same events. (a) transport plan between a $\pi^0$ event (red) and an ${e}^-$ event (blue) (b) transport plan between the same ${e}^-$ event (blue) and another ${e}^-$ event (turquoise).}
    \label{fig:2D-Transport-Plans}

\end{figure*}

\subsection{Projections}
Past applications of OT in collider physics~\cite{Cai:2020vzx, Cai:2021hnn} have largely treated calorimeter deposits as distributions on the two-dimensional (2D) ground space comprising the longitudinal and azimuthal plane of the calorimeters in collider detectors. 
This 2D representation of calorimeter deposits at colliders is physically well-motivated as much of the relevant information in the underlying event is captured by the projection of outgoing particles onto the celestial sphere. In contrast, the reconstruction of neutrino interactions enabled by the 3D calorimetry in LArTPC detectors motivates treating calorimeter deposits in neutrino experiments as distributions on the full 3D ground space, as shown in Figure~\ref{fig:3D-Transport-Plans}.
This presumes that the full 3D distributions contain information relevant for event classification. 

However, since showers tend to leave small residual charge deposits at a large distance from their shower start, (see Fig.~\ref{fig:3DElectron}) the OT transport plan may be negatively impacted by the large cost to transport this residual charge deposited far along the shower axis. These large distances are not representative of differences between an $e$ and $\pi^0$ topology, and may lead to sub-optimal performance, particularly when relying on the full 3D calorimetric information available from the detector.

Hence, inspired by event classification for jets at the LHC, we also explore using 2D projections along the principal axis component, which is analogous to the so-called jet images in colliders. The primary similarity here is in the construction of a physically relevant axis along which to take the projection such that angular deviations carry physical meaning. Figure.~\ref{fig:2D-Transport-Plans} shows a comparison of OT plans for an $e$-$\pi^0$ pair (left) and two electrons (right) performed both in 3D (insets in each figure) as well as within the 2D charge projection (main panel). The particularly long distance covered by a small component of the electron in cyan along the vertical axis (inset of Fig.~\ref{fig:2DEE-Transport}) causes a 3D OT distance which is much larger than the 3D OT distance for the corresponding $e$-$\pi^0$ pair (Fig.~\ref{fig:2DPE-Transport}). The 2D projected charge mitigates this, leading (in this specific example) to a much smaller computed 2D OT distance for the $e$-$e$ pair compared to the $e$-$\pi^0$ pair.

\section{Methods}\label{sec:methods}

Optimal transport distance computations are carried out using the Python Optimal Transport (POT) library \cite{flamary_2021}, which provides several generic OT solvers. In order to have a balanced sample, the same number of $\pi^0$ and $e$ events are chosen for each energy bin so we have $N_{\pi^0}=N_e$. 
To allow for adequate sample statistics in each bin, the bin boundaries are chosen to enforce $N_{e,p}$ to each be greater than 1,000. The bin edges are defined as [0.05, 0.1, 0.15, 0.2, 0.25, 0.3, 0.35, 0.4, 0.5, 0.9] GeV. 
For each bin, these $N_{\pi^0}+N_e$ events are divided into two equally sized training and testing sets, $N_{\pi^0,train}+N_{e,train}$ and $N_{\pi^0,test}+N_{e,test}$, respectively. 

Given a test event, of unknown type, each OT method must assign a single number (score) which indicates the likely identity of  that test event. The exact details of how the score is derived depend on the particular strategy employed. We explore ``Minimum OT'', k-Nearest Neighbors, and Support Vector Machine and describe each below.

\subsection{Minimum OT: min(OT)}
The first method we employ takes advantage of the fact that the OT distance between like-distributions will be smaller than the distance between unlike-distributions. Since we expect physically similar events to have similar detector signatures, we can use this property to our advantage. In this case, the OT distance between a $\pi^0$ event and electron event will generally be larger than the distance between two electron events. 

In the min(OT) method, the strategy employed to identify $\pi^0$ events is to first designate a number of electron events from the training set as a reference, $\{ \mathcal{E}_{e_i} \mid i = 1, ..., N_{e,ref}\}$. We can then compare an event of an unknown type, $\mathcal{T}$, to this reference population. Namely, for each event $\mathcal{T}$ we can calculate $N_{e,ref}$ OT distances $\{ {\rm OT}(\mathcal{T}, \mathcal{E}_{e_i}) ~\mid~ i=1,..., N_{e,ref} \}$. Electron events, $\mathcal{T}_e$, will tend to have a smaller OT distance than $\pi^0$ events, $\mathcal{T}_{\pi^0}$ i.e. the set $\{{\rm OT}(\mathcal{T}_e, \mathcal{E}_{e_i})\}$ will generally be closer to 0 than $\{{\rm OT}(\mathcal{T}_{\pi^0}, \mathcal{E}_{e_i})\}$. The task now is to turn $\{ {\rm OT}(\mathcal{T}, \mathcal{E}_{e_i}) ~\mid~ i=1,..., N_{e,ref} \}$ into a single score for each test event $\mathcal{T}$.

We choose to take the minimum of these distances as our score for that test event. Namely, the score for an event $\mathcal{T}$ is the OT distance to the closest electron event in the reference sample.
This approach follows from previous work  on LHC data~\cite{Craig:2024rlv} which explored several methods of reduction. We also note that there have been several works in the context of particle colliders which have argued that the notion of smallest OT distance is a valid discriminating observable~\cite{Komiske:2020qhg, Fraser:2021lxm, Craig:2024rlv}. Ref.~\cite{Komiske:2020qhg} even found that in some cases  traditional physics observables correspond to the closest OT distance to a manifold of events.

\begin{figure}
   \includegraphics[width=0.9\columnwidth]{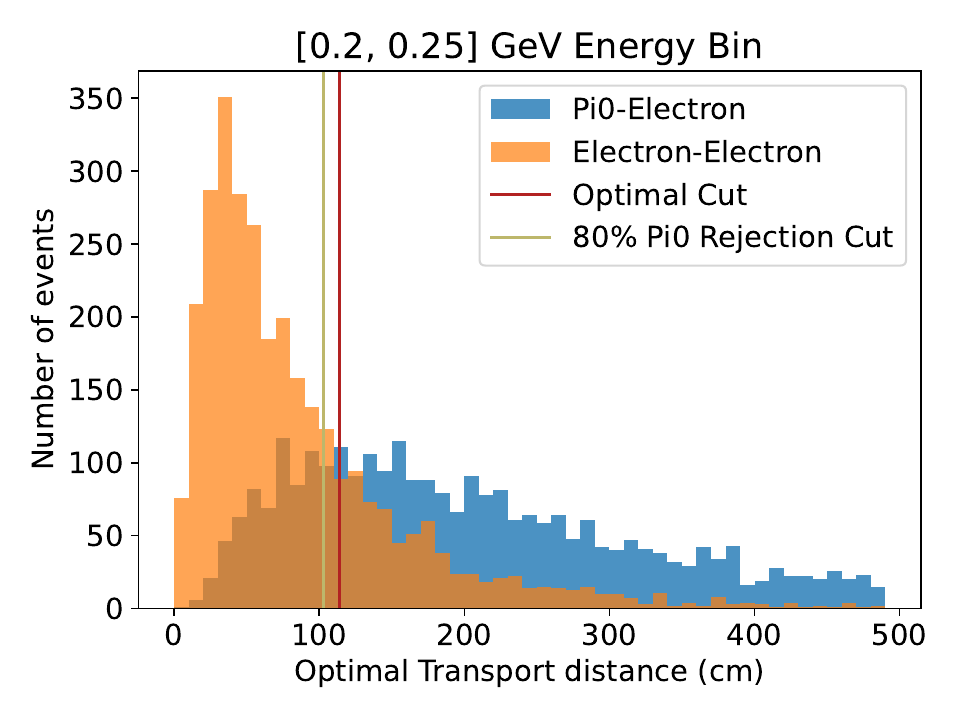}
    \caption{Histogram of the minimum OT distance between $\pi^0$ and $e$ events (blue) and between different electron events (orange). Two different cut thresholds for min(OT) values are shown; one gives the highest accuracy (red) and the other achieves $80\%$ $\pi^0$ rejection (yellow).}
    \label{fig:OT-Hist}
\end{figure}

Figure~\ref{fig:OT-Hist} shows the distribution of min(OT) scores calculated for the electron and $\pi^0$ test samples in the 0.2-0.25 GeV energy bin. As expected, the OT distance between like-events (electron-electron) tends to be smaller than the OT distance between unlike-events ($\pi^0$-electron).
Two different thresholds for min(OT) values are shown representing cuts which give the highest accuracy (red) or a representative $80\%$ $\pi^0$ rejection (gold).

We reserve half of the  electron events to act as a reference sample and the other half are reserved for testing; the same number of $\pi^0$ events similarly are selected for testing.  To get uncertainty estimates on our results, we further split our testing sample containing both $\pi^0$s and electrons into 5 randomly shuffled sets each of size $N_{test}$. The min(OT) distance between each test event and the $N_{e,ref}$ electrons is computed. We can then obtain ROC curves and resulting metrics (e.g. accuracy) from various threshold cuts on min(OT) scores. We repeat this process for each of the 5 test sets and report the average and standard deviations of the results.

\subsection{Post-processing pairwise OT distances with Machine Learning}
Alternatively, interpretable ML methods can be used on pairwise OT distances to further boost the classification performance. The two methods we consider are k-nearest neighbors (kNN) and support vector machine (SVM). 
These methods overcome the fact that while two distributions may not be perfectly separated, OT distances for like-events will tend to cluster together, providing additional information that can be used for event classification. These methods have also been shown to boost performance in certain contexts~\cite{Cai:2020vzx,Craig:2024rlv}. 

In contrast, the min(OT) method relies on the distribution of $\min \{ {\rm OT}(\mathcal{T}_{\pi^0}, \mathcal{E}_{e_1}), ..., {\rm OT}(\mathcal{T}_{\pi^0}, \mathcal{E}_{e_{N_{e,ref}}}) \}$ being sufficiently separated from the distribution of $\min \{ {\rm OT}(\mathcal{T}_{e}, \mathcal{E}_{e_1}), ..., {\rm OT}(\mathcal{T}_{e}, \mathcal{E}_{e_{N_{e,ref}}}) \}$. While this allows you to operate in an unsupervised manner, you risk sacrificing additional separation power that supervised post-processing (kNN and SVM) can yield. However, these methods now require both electron and $\pi^0$ events for training. Namely, the inputs to these models are the full set of pairwise OT distances, $\{ {\rm OT} (\mathcal{E}_{\pi^0}, \mathcal{E}_{e}) \}$, for all $e, \pi^0$ training events.

\subsubsection{K-Nearest Neighbors (kNN)}\label{sec:kNN}
k-Nearest Neighbors (kNN)~\cite{kNN_1967} is a supervised algorithm which can be used for classification tasks, where the hyperparameter $k$ is the number of nearest neighbors. 
An $n$-dimensional OT distance space is constructed, where $n$ is the number of events in the training sample. 
kNN makes a prediction for a test event by finding the $k$ events in the training sample which are closest to the test event we are trying to classify. Each neighbor casts a vote corresponding to its own type; the majority type is then assigned to the test event and a percentage given by the fraction of neighbors of that type is assigned as a measure of the probability that the test event is of that type. For example, if all $k$ neighbors are electrons, the test event will be assigned the particle type `electron' with 100\% probability.

To estimate the best hyperparameter values, we implement 5-fold cross-validation on the randomly mixed training dataset $N=N_{\pi^0,train}+N_{e,train}=500$.
We first fix a hyperparameter value $k$.
We split $N$ into five separate sets, $N^i$ for $i=1,...,5$. We select one of these to be our validation set and the remaining four to be our training e.g. $N_{val}=N^0=100$ and $n=\sum_{j\neq i} N^j = 400$. We store the results of the kNN, then repeat the process until we have cycled through all possible choices of $N_{val}=N^i$.  We then average the resulting 5 sets of stored kNN results.
We repeat this process for a range of hyperparameter values $k$, and the best performing hyperparameter is selected.
Initial scans of the hyperparameter $k$ shows best performance around $k=9$, which is why for all energy bins we scan over the hyperparameter space in the range $k \in [7, 11]$ with an increment size of $1$.
Similar to what was done with min(OT), the final results are obtained by splitting the $N_{\pi^0,test}+N_{e,test}$ into 5 sets of size 100. The model with the best hyperparameter $k$ is retrained using the full $N_{\pi^0,train}+N_{e,train}$ set of events and is then evaluated on the 5 test sets; the average and standard deviation of the results are reported.

\subsubsection{Support Vector Machine (SVM)}
Similarly, a Support Vector Machine (SVM)~\cite{SVM_1999} can also be used as a supervised classifier.
Events are again distributed in a hyperspace constructed from OT distances. However, instead of majority vote by the neighbors, a hyperplane (called the decision boundary) which most effectively separates the two classes is found. A prediction for the class of an event is made based on which side of the hyperplane an event lies.

There are two hyperparameters in SVM:
\begin{itemize}
    \item $C$, regularization parameter controlling the margin size around the decision boundary
    \item $\gamma$, which determines the effect of a training event on the decision boundary
\end{itemize}
In particular, we choose a Gaussian RBF kernel in which $\gamma$ can be interpreted as the inverse variance of the kernel. A larger value of $\gamma$ means training events which are far away from the decision boundary in OT-distance space are weighted less when constructing the decision boundary.
In the case of using OT distances for $e/\pi^0$ classification, we found that the value of $C$ had little effect on the results. 
As for $\gamma$, we first performed a coarse scan to determine the best $\gamma$ range of $[0.002, 0.014]$. We then carried out a finer scan with an increment of $0.002$ to find the optimal choice of $\gamma$.
As in Sec.~\ref{sec:kNN}, 5-fold cross validation is performed to select the best hyperparameters. And the average and standard deviation of the results over 5 test sets are reported.

\section{\label{sec:results} Results}

\begin{figure*}[ht!]
\makebox[\textwidth][c]{
\includegraphics[width=1.25\textwidth]{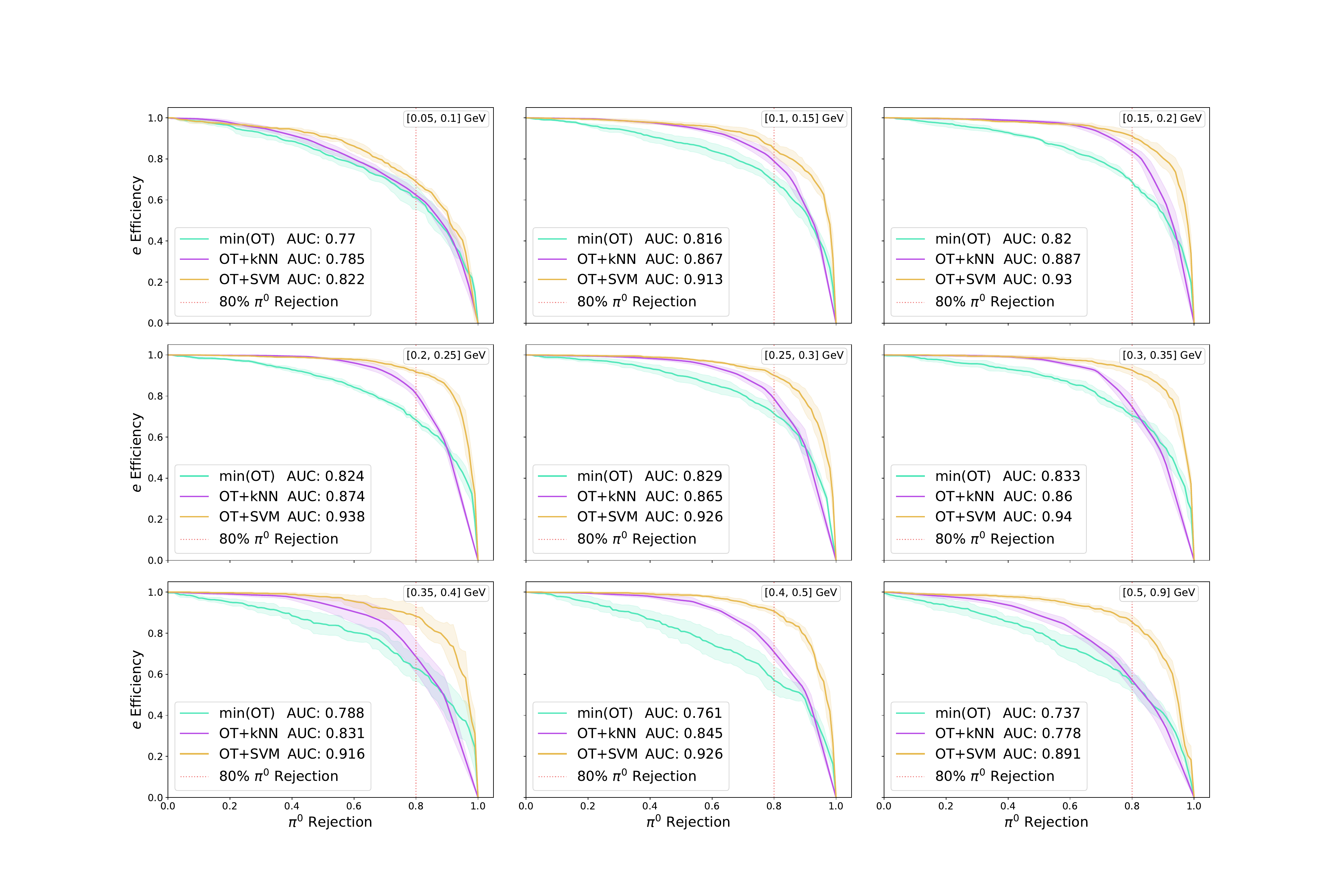}
}
\caption{ROC curves for 3D OT methods between events in each energy bin (GeV).}
\label{fig:ROC}
\end{figure*}

In this section, we show the performance of the methods (min(OT), OT+kNN, and OT+SVM) described in Sec.~\ref{sec:methods}.  Table \ref{table:results} shows the performance of these three methods in each of the 9 energy bins.  The performance metrics we consider are: the area under the curve (AUC), accuracy of overall classification, and $e$ efficiency at 80\% $\pi^0$ rejection. The receiver operating characteristic (ROC) curves for the 9 energy bins are shown in Figure~\ref{fig:ROC}.

Overall, the best performing method has an accuracy ranging from 0.828 to 0.876 for events above 100 MeV. Performance is shown to be relatively stable across all energy bins, with a slight drop at higher energies. 
The first energy bin at $0.05-0.1$ GeV is also shown to have lower relative performance. This is to be expected since this energy range is below the rest mass of a $\pi^0$, meaning the $\pi^0$ events in this energy bin are missing a significant portion of their energy due to effects such as inefficiencies in the 3D reconstruction and a fraction of the charge escaping the detector boundary. 
Overall, all methods perform well, however post-processing with both kNN and SVM boosts performance. SVM in particular makes the greatest improvement. In-line with previous works~\cite{Cai:2020vzx, Cai:2021hnn}, interpretable ML methods are shown to make good use of the features of OT distances and yield more competitive performance.
Using 2D projections as input for OT (abbreviated as 2DOT) is shown to perform better than 3DOT when using min(OT) as the post-processing method, which makes sense as min(OT) may be particularly sensitive to cases such as those discussed in Fig.~\ref{fig:2D-Transport-Plans}. However, kNN and SVM are able to overcome this difference. In particular, 3DOT+SVM has the best performance across all energy bins.
As we will see in Sec.~\ref{sec:performanceCFpandora}, all three methods with both 2D and 3D OT outperform traditional reconstruction in $e/\pi^0$ classification.

To contextualize these results, it is important to note that in a full analysis where only one photon can be clearly identified, other handles for $e$/$\gamma$ separation would be subsequently employed to remove remaining backgrounds. 
Handles such as calorimetric d$E$/d$x$, or the presence of a ``gap'' allow for additional discriminating power. Importantly, however, the more $e$/$\pi^0$ classification can be improved by correctly identifying the characteristic two-shower vs. one-shower topology of the event early in the classification, the less one needs to rely on subsequent cuts such as d$E$/d$x$, the presence of a ``gap'', or other kinematic-based discriminating variables. These subsequent cuts may introduce several drawbacks including larger inefficiencies and potential kinematic and topological biases. Boosting $e$/$\pi^0$ separation through the topological discrimination of the full energy deposit as OT is able to do can therefore lead to a meaningful performance gains when applied to physics measurements in LArTPCs.

\begin{table*}[p]
\centering
\renewcommand{\arraystretch}{1.2}
\resizebox{\textwidth}{!}{\begin{tabular}{c|c|c|c|c|c|c|c}
E Bin  & Performance    &           \multicolumn{3}{c|}{2DOT}& 
          \multicolumn{3}{c}{3DOT}\\ \hline
(GeV)        & Metric         & min(OT)   & kNN          & SVM  & min(OT)   & kNN          & SVM  \\ \hline
                  & AUC              & $0.786 \pm 0.017$ & $0.761 \pm 0.013$ & $0.783 \pm 0.016$ & $0.770 \pm 0.022$& $0.785 \pm 0.007$& $\mathbf{0.822 \pm 0.007}$\\
0.05-0.1          & accuracy         & $0.738 \pm 0.022$ & $0.699 \pm 0.016$ & $0.717 \pm 0.019$ & $0.717 \pm 0.020$& $0.694 \pm 0.008$& $\mathbf{0.742 \pm 0.012}$\\
                  & $e^-$ efficiency & $0.649 \pm 0.052$ & $0.574 \pm 0.044$ & $0.612 \pm 0.026$ & $0.611 \pm 0.058$& $0.624 \pm 0.023$& $\mathbf{0.689 \pm 0.022}$\\ \hline
                  & AUC              & $0.826 \pm 0.021$ & $0.852 \pm 0.015$ & $0.852 \pm 0.015$ & $0.816 \pm 0.024$& $0.867 \pm 0.007$& $\mathbf{0.913 \pm 0.009}$\\
0.1-0.15          & accuracy         & $0.765 \pm 0.022$ & $0.780 \pm 0.023$ & $0.791 \pm 0.018$ & $0.755 \pm 0.015$& $0.784 \pm 0.009$& $\mathbf{0.828 \pm 0.024}$\\
                  & $e^-$ efficiency & $0.712 \pm 0.047$ & $0.754 \pm 0.044$ & $0.780 \pm 0.043$ & $0.693 \pm 0.034$& $0.788 \pm 0.032$& $\mathbf{0.848 \pm 0.036}$\\ \hline
                  & AUC              & $0.836 \pm 0.007$ & $0.850 \pm 0.012$ & $0.856 \pm 0.008$ & $0.820 \pm 0.008$& $0.887 \pm 0.016$& $\mathbf{0.930 \pm 0.015}$\\
0.15-0.2          & accuracy         & $0.765 \pm 0.009$ & $0.774 \pm 0.019$ & $0.792 \pm 0.007$ & $0.754 \pm 0.009$& $0.803 \pm 0.019$& $\mathbf{0.858 \pm 0.012}$\\
                  & $e^-$ efficiency & $0.718 \pm 0.018$ & $0.752 \pm 0.025$ & $0.789 \pm 0.013$ & $0.688 \pm 0.006$& $0.837 \pm 0.023$& $\mathbf{0.910 \pm 0.025}$\\ \hline
                  & AUC              & $0.839 \pm 0.007$ & $0.856 \pm 0.018$ & $0.860 \pm 0.018$ & $0.824 \pm 0.009$& $0.874 \pm 0.009$& $\mathbf{0.938 \pm 0.010}$\\
0.2-0.25          & accuracy         & $0.767 \pm 0.006$ & $0.782 \pm 0.017$ & $0.788 \pm 0.014$ & $0.751 \pm 0.007$& $0.788 \pm 0.021$& $\mathbf{0.876 \pm 0.008}$\\
                  & $e^-$ efficiency & $0.720 \pm 0.015$ & $0.755 \pm 0.050$ & $0.777 \pm 0.034$ & $0.683 \pm 0.014$& $0.811 \pm 0.027$& $\mathbf{0.917 \pm 0.015}$\\ \hline
                  & AUC              & $0.847 \pm 0.013$ & $0.850 \pm 0.014$ & $0.852 \pm 0.015$ & $0.829 \pm 0.019$& $0.865 \pm 0.014$& $\mathbf{0.926 \pm 0.010}$\\
0.25-0.3          & accuracy         & $0.774 \pm 0.013$ & $0.776 \pm 0.014$ & $0.784 \pm 0.012$ & $0.765 \pm 0.013$& $0.768 \pm 0.012$& $\mathbf{0.853 \pm 0.010}$\\
                  & $e^-$ efficiency & $0.734 \pm 0.026$ & $0.768 \pm 0.024$ & $0.760 \pm 0.018$ & $0.714 \pm 0.033$& $0.787 \pm 0.042$& $\mathbf{0.900 \pm 0.020}$\\ \hline
                  & AUC              & $0.853 \pm 0.022$ & $0.862 \pm 0.020$ & $0.879 \pm 0.030$ & $0.833 \pm 0.019$& $0.860 \pm 0.013$& $\mathbf{0.940 \pm 0.012}$\\
 0.3-0.35         & accuracy         & $0.786 \pm 0.023$ & $0.797 \pm 0.028$ & $0.815 \pm 0.013$ & $0.762 \pm 0.010$& $0.778 \pm 0.016$& $\mathbf{0.873 \pm 0.018}$\\
                  & $e^-$ efficiency & $0.749 \pm 0.063$ & $0.783 \pm 0.053$ & $0.828 \pm 0.024$ & $0.703 \pm 0.025$& $0.747 \pm 0.049$& $\mathbf{0.925 \pm 0.029}$\\ \hline
                  & AUC              & $0.818 \pm 0.023$ & $0.834 \pm 0.035$ & $0.848 \pm 0.040$ & $0.788 \pm 0.034$& $0.831 \pm 0.035$& $\mathbf{0.916 \pm 0.023}$\\
 0.35-0.4         & accuracy         & $0.758 \pm 0.024$ & $0.759 \pm 0.027$ & $0.765 \pm 0.035$ & $0.732 \pm 0.027$& $0.710 \pm 0.022$& $\mathbf{0.837 \pm 0.032}$\\
                  & $e^-$ efficiency & $0.686 \pm 0.047$ & $0.705 \pm 0.066$ & $0.751 \pm 0.080$ & $0.628 \pm 0.063$& $0.685 \pm 0.079$& $\mathbf{0.881 \pm 0.047}$\\ \hline
                  & AUC              & $0.822 \pm 0.016$ & $0.851 \pm 0.015$ & $0.857 \pm 0.014$ & $0.761 \pm 0.035$& $0.845 \pm 0.009$& $\mathbf{0.927 \pm 0.006}$\\
 0.4-0.5          & accuracy         & $0.756 \pm 0.014$ & $0.777 \pm 0.022$ & $0.787 \pm 0.011$ & $0.711 \pm 0.028$& $0.741 \pm 0.013$& $\mathbf{0.857 \pm 0.012}$\\
                  & $e^-$ efficiency & $0.688 \pm 0.026$ & $0.742 \pm 0.025$ & $0.760 \pm 0.027$ & $0.57 \pm 0.068$& $0.708 \pm 0.036$& $\mathbf{0.907 \pm 0.020}$\\ \hline
                  & AUC              & $0.783 \pm 0.033$ & $0.822 \pm 0.015$ & $0.833 \pm 0.022$ & $0.737 \pm 0.034$& $0.778 \pm 0.015$& $\mathbf{0.891 \pm 0.014}$\\
 0.5-0.9          & accuracy         & $0.733 \pm 0.037$ & $0.751 \pm 0.021$ & $0.763 \pm 0.026$ & $0.689 \pm 0.031$& $0.674 \pm 0.029$& $\mathbf{0.828 \pm 0.016}$\\
                  & $e^-$ efficiency & $0.646 \pm 0.073$ & $0.694 \pm 0.044$ & $0.738 \pm 0.044$ & $0.553 \pm 0.044$& $0.571 \pm 0.045$& $\mathbf{0.854 \pm 0.033}$\\             
\end{tabular}}
\caption{Optimal transport performance for min(OT), kNN and SVM combined with 2DOT and 3DOT in 9 energy bins. Performance metrics presented here include: area under the curve (AUC), accuracy of overall classification and $e$ efficiency at 80\% $\pi^0$ rejection.}
\label{table:results}
\end{table*}

\subsection{Performance Compared with Pandora}\label{sec:performanceCFpandora}
In order to benchmark results with current state-of-the-art tools,  performance of OT is compared to the Pandora reconstruction framework~\cite{bib:uBpandora,bib:protoDUNEpandora} as implemented in the reconstruction available in MicroBooNE's Open Datasets. Pandora is chosen because it is currently the most widely used reconstruction framework across neutrino LArTPC experiments, and because its reconstruction output is automatically stored in the MicroBooNE open dataset files used for this work. In comparing the performance of OT to a more traditional reconstruction approach, we use the number of reconstructed EM showers in an event as a way to classify electrons (one shower) vs. $\pi^0$ (two showers). We find that overall OT outperforms Pandora in all phase space for certain kinematic variables.
In this comparison it is important to keep in mind that the pre-processing employed for this work (Sec.~\ref{sec:data_processing}) relies on truth-labeling of EM shower activity which is not available from the already stored Pandora output. The impact this has in the performance comparisons presented was tested by studying based on truth information how classification results change as the quality of the reconstruction increases. We find that overall this does not meaningfully alter the conclusions of this comparison. As mentioned previously, future integration of OT in event classification would be paired with existing complementary reconstruction packages that can perform hit-level track-shower classification with high accuracy, such as NuGraph2~\cite{bib:NuGraph}. This would remove the reliance on truth-based information leveraged for this study.

\subsubsection{$\pi^0$ Classification}

To successfully classify an event as a $\pi^0$ in Pandora, we require that two or more reconstructed showers are present in the event. For OT, we rely on the methods described earlier. The reason we are requiring two or more showers instead of exactly two showers is to aim for fairly evaluating Pandora's performance at distinguishing electrons from $\pi^0$s without demanding stricter reconstruction accuracy for the individual photon showers in the event.

\begin{figure}[hbt]
    \centering
   \includegraphics[width=0.9\columnwidth]{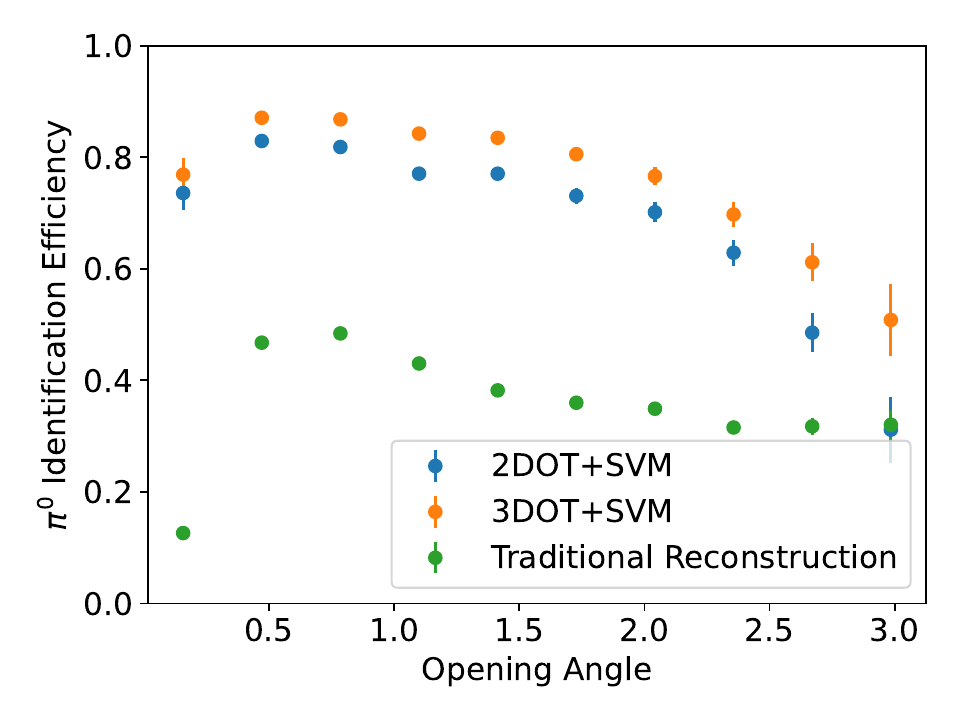}
    \caption{Performance compared with opening angle. For Pandora, $\pi^0$ efficiency is defined as its efficiency in finding $\ge 2$ showers.  Error bars represent statistical uncertainties of the samples in each bin.}
    \label{fig:opening-angle}
\end{figure}

\begin{figure}[hbt]
    \centering
   \includegraphics[width=0.9\columnwidth]{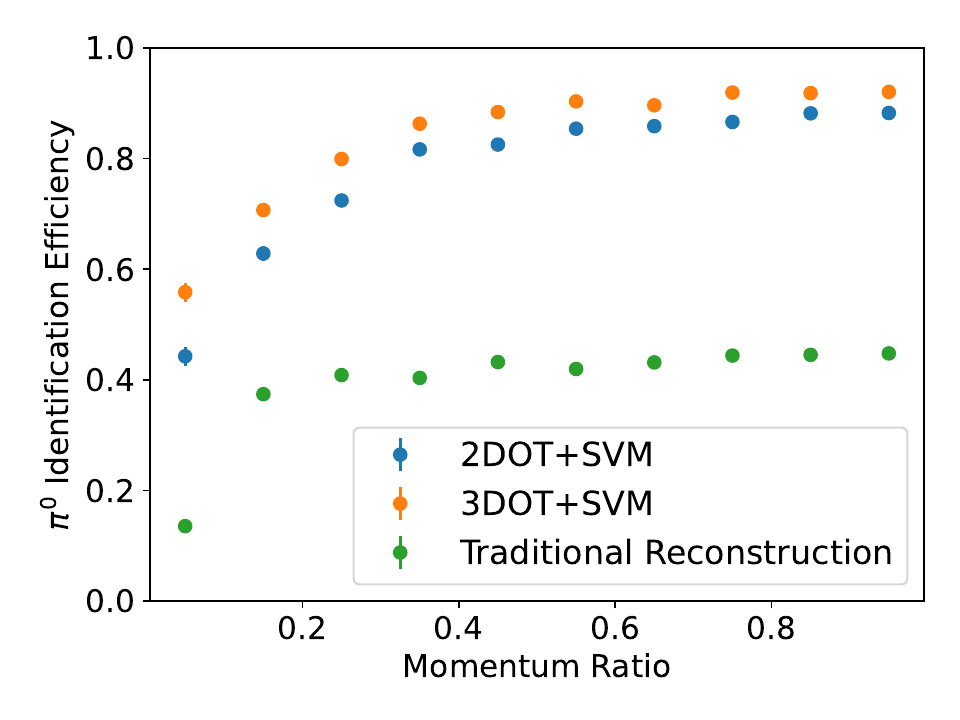}
    \caption{Performance compared with subleading shower momentum ratio. For Pandora, $\pi^0$ efficiency is defined as its efficiency in finding $\ge 2$ showers. Error bars represent statistical uncertainties of the samples in each bin.}
    \label{fig:subleadp-ratio}
\end{figure}

\begin{figure*}[p]
\subfloat[\label{fig:SOA-P157E44}]{%
  \includegraphics[width=.34\textwidth]{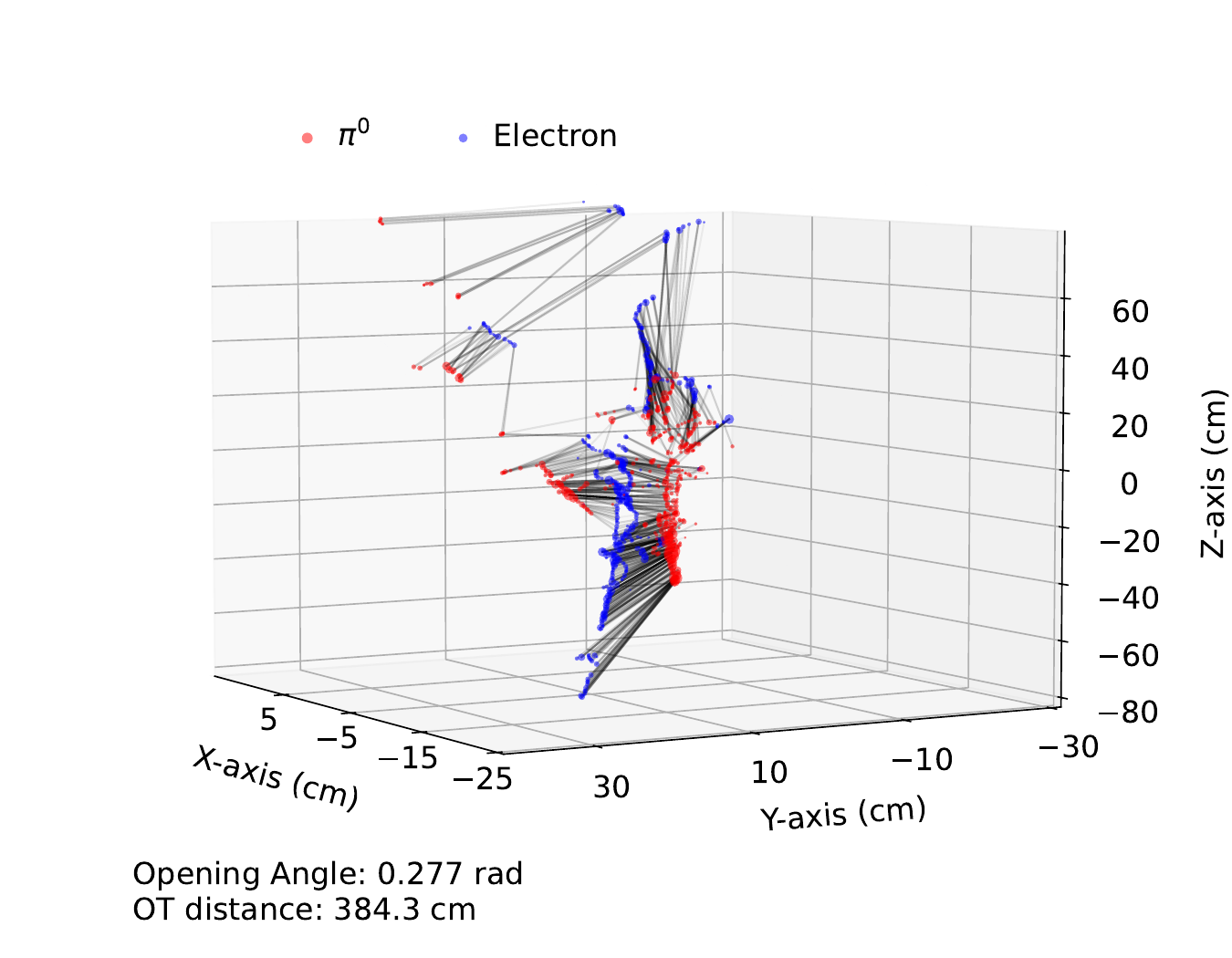}  
}\hspace*{\fill}%
\subfloat[\label{fig:SOA-P158E216}]{%
  \includegraphics[width=.34\textwidth]{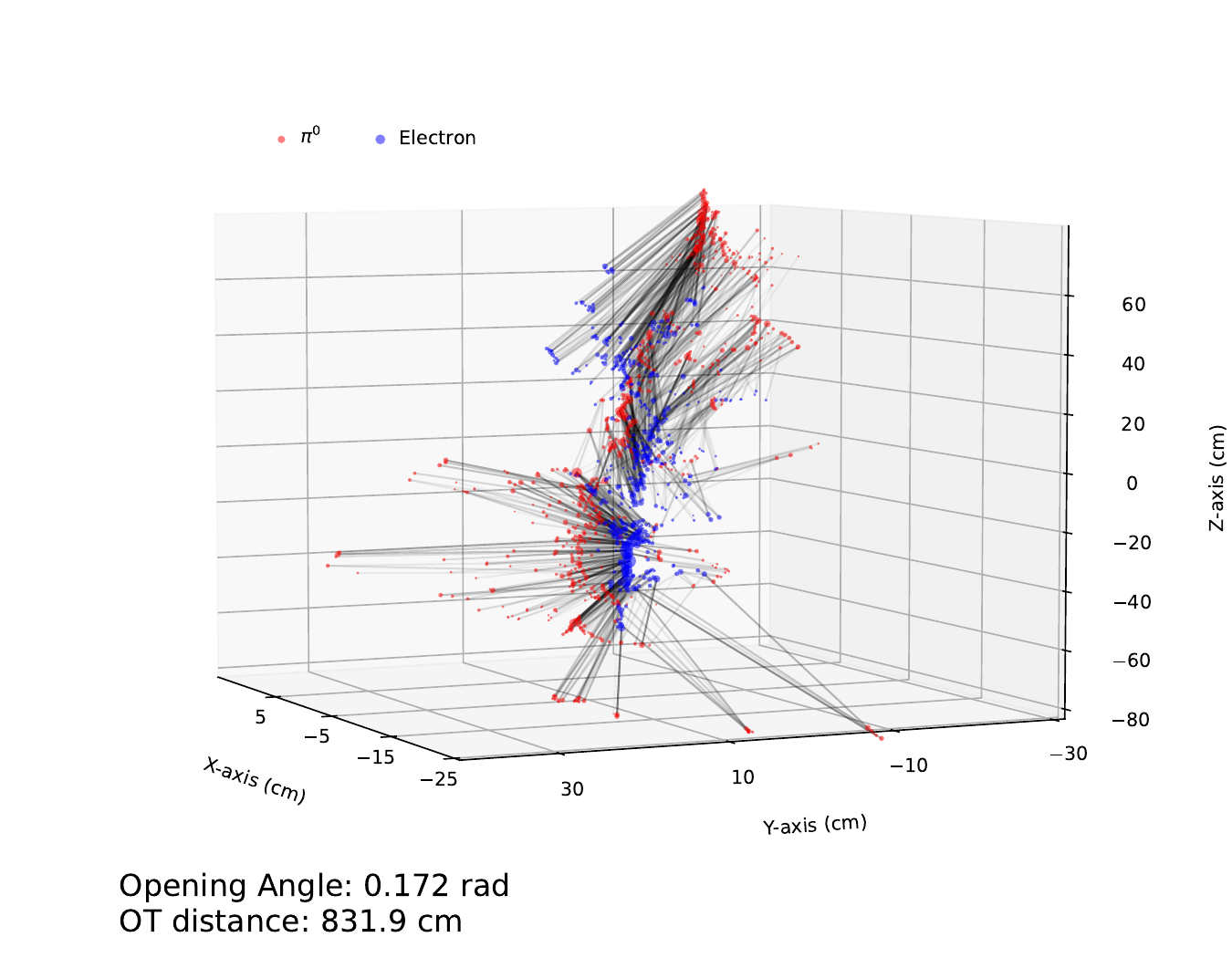}  
}\hspace*{\fill}%
\subfloat[\label{fig:SOA-P21E61}]{%
  \includegraphics[width=.34\textwidth]{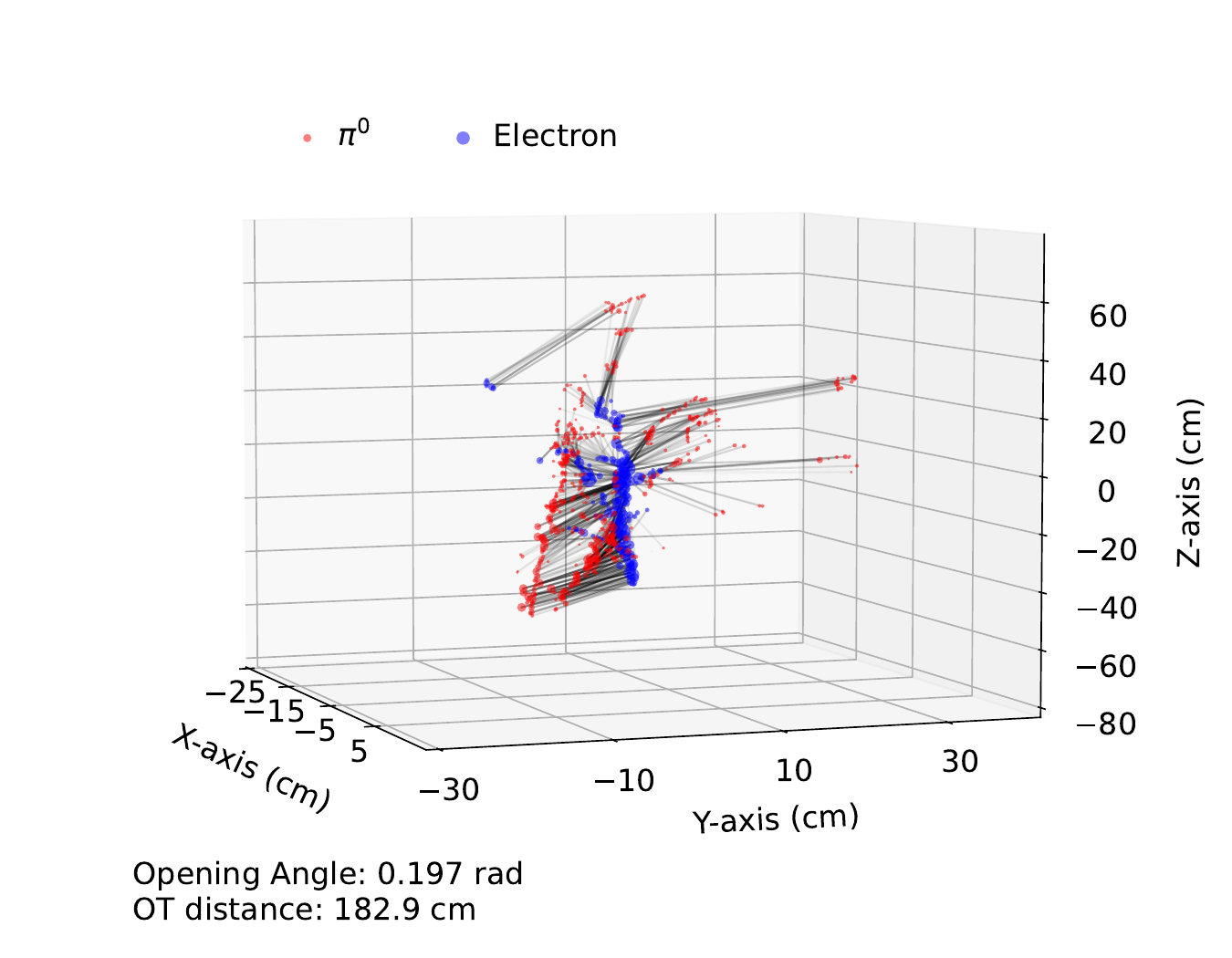}  
}
\caption{Transport plans between $\pi^0$ with small opening angle and electrons in the same energy bin.}
\label{fig:SmallOpeningAngle}
\end{figure*}

The results are presented as a function of the relevant kinematic variables for $\pi^0 \rightarrow \gamma\gamma$ decays. The first variable studied is the momentum ratio of the two photons, defined as $R = E_{\gamma 2} / E_{\gamma 1}$ where $E_{\gamma 1}$ and $E_{\gamma 2}$ are the energy of the leading and sub-leading photons, respectively. Events with low momentum ratio indicate a very low-energy, sub-leading photon, which is typically challenging to identify. The second variable used is the opening angle between the  photons produced by the $\pi^0$ decay. Photon showers which are closely aligned with a small opening angle are another typically challenging topology to identify and reconstruct. The results are shown in Figures~\ref{fig:opening-angle} and~\ref{fig:subleadp-ratio}, and compare Pandora's performance (green) to that of OT using 3D (orange) and 2D (blue) projections with SVM post-processing .
Both 2DOT and 3DOT have more competitive performance in $\pi^0$ classification. As compared to Pandora which has an overall efficiency of 40\%, OT is able to correctly classify 80\% of the $\pi^0$s in general. The improvement is particularly pronounced for events with small opening angles, where Pandora is only able to correctly classify $\sim10\%$ of $\pi^0$s (see Figure \ref{fig:opening-angle}).
The three different examples of  $\pi^0$ events with small opening angles in Fig.~\ref{fig:SmallOpeningAngle} show why this could be the case. In Figures \ref{fig:SOA-P157E44} and \ref{fig:SOA-P158E216} the electron showers are spatially located between the two $\gamma$ showers. In Figure \ref{fig:SOA-P21E61}, the electron shower overlaps with one photon shower while being at a distance away from the other one. In all three cases, we see that much work was done to overcome the planar distributional differences between one-shower and two-shower structures, despite the small opening angles for the $\pi^0$ events, leading to good event classification even in this challenging region of phase space.
When evaluating the performance as a function of the momentum ratio (Fig.~\ref{fig:subleadp-ratio}), we similarly notice a particularly large performance improvement at low values of momentum ratio, improving the classification efficiency from around 15\% to almost 60\% for the lowest bin.

\subsubsection{Electron Classification}
\begin{figure}[hbt]
    \centering
   \includegraphics[width=0.9\linewidth]{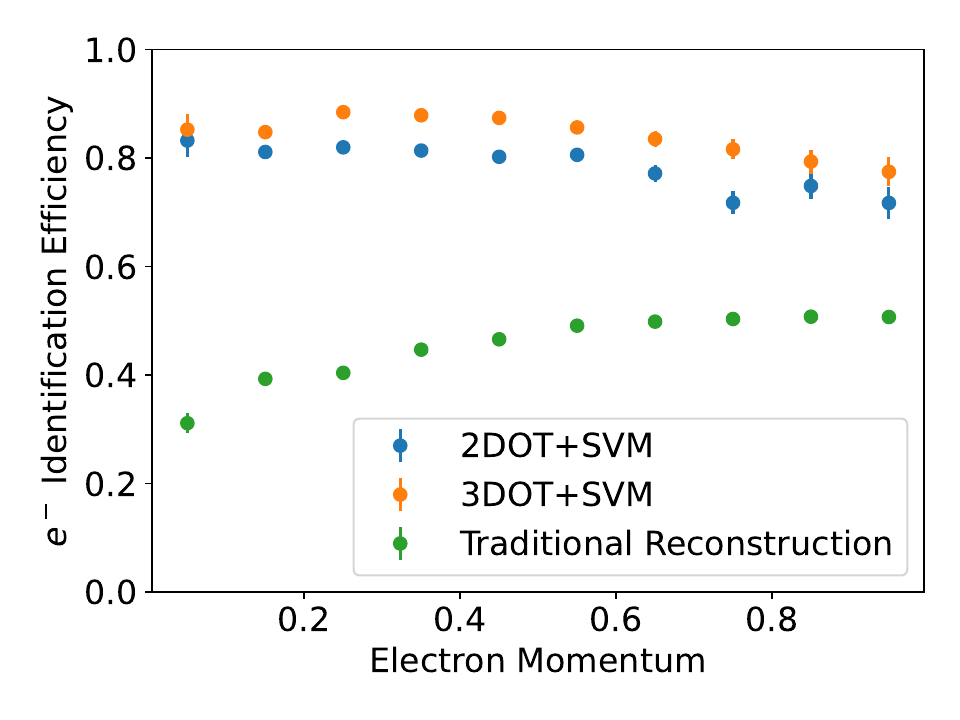}
    \caption{Performance as a function of electron momentum. For Pandora, $e^-$ efficiency is defined as its
efficiency in reconstructing one shower in the event.}
    \label{fig:electron-momentum}
\end{figure}

We additionally evaluate the performance as a function of the electron momentum, as shown in Fig.~\ref{fig:electron-momentum}.  Here too, we see that OT outperforms traditional reconstruction across all values of electron momentum, with a relatively stable performance  around 80\%, as opposed to Pandora which has an overall efficiency of about 40\%. 

\subsection{Computational Scaling of OT Calculation}\label{sec:computationEfficiency}

In this work, we have used a standard strategy to calculate OT distances~\cite{Bonneel2011} which is implemented in the POT library~\cite{flamary_2021}. The average CPU time for a single OT distance computation for each energy bin is shown in Table \ref{table: CPUTime}. By single OT distance computation we mean calculating the OT distance between two events, ${\rm OT}(\mathcal{E}, \mathcal{E}').$
However, more efficient OT distance strategies exist. In this section, we briefly review the computational time complexity of OT distance calculations in general and in the context of this work. Our hope in doing so is to motivate the wide array of OT approximation methods that could be utilized to deploy these methods at scale in LArTPC experiments.

\begin{table}[hbt!]
\centering
\renewcommand{\arraystretch}{1.2}
\begin{tabular}{c|c|c}
E Bins (GeV) & 2DOT(ms) & 3DOT(ms)\\ \hline
0.05-0.1          & $0.364 \pm 0.015$  &$0.382 \pm 0.024$\\
0.1-0.15          & $0.629 \pm 0.012$  &$0.624 \pm 0.012$\\
0.15-0.2          & $0.986 \pm 0.033$  &$0.963 \pm 0.028$\\
0.2-0.25          & $1.46 \pm 0.02$    &$1.62 \pm 0.48$\\
0.25-0.3          & $2.18 \pm 0.37$    &$2.22 \pm 0.69$\\
0.3-0.35          & $2.32 \pm 0.51$    &$2.58 \pm 0.052$\\
0.35-0.4          & $3.26 \pm 0.75$    &$3.68 \pm 0.35$\\
0.4-0.5           & $8.24 \pm 0.56$    &$8.78 \pm 0.88$\\
0.5-0.9           & $29.7 \pm 1.9 $    &$27.8 \pm 3.3$
\end{tabular}
\caption{CPU time for a single OT distance computation in $\mathrm{ms}$. These times were averaged over $1,000$ computations on a standard laptop. The average and standard deviation of the time is reported.}
\label{table: CPUTime}
\end{table}

The time complexity of an OT distance calculation is dominated by the dependence on the number of discrete probability masses, $n$. In the most pessimistic case, the computational complexity of OT distance calculations grows as $\mathcal{O}(n^3~{\rm log}~n)$~\cite{Cuturi2013}. However, the algorithm implemented in  Ref.~\cite{Bonneel2011} lowers this to $\mathcal{O}(n^3)$. Additionally, many approximate algorithms exist which are even more efficient. For example, Sinkhorn distances~\cite{cuturi2013sinkhorndistanceslightspeedcomputation} scale as $\mathcal{O}(n^2)$\footnote{Although more recent variants claim near linear performance~\cite{altschuler2018}.} with a sample-complexity of $\mathcal{O}(n^{-1/2})$~\cite{genevay2019} and sliced Wasserstein distances scale as $\mathcal{O}(n~ {\rm log}~n)$ also with a sample-complexity of $\mathcal{O}(n^{-1/2})$~\cite{nguyen2023}. Interestingly,  neither of these strategies suffer from the curse of dimensionality and thus still work efficiently when the dimension $d$ of the underlying metric space is large.

In addition to the time complexity due to $n$, we must also consider how many OT distance calculations are required. For a single test event and $N_{\rm ref}$ reference events, we must calculate $N_{\rm ref}$ OT distances. Thus, the overall complexity per test event is at worst $\mathcal{O}(N_{\rm ref}~n^3~{\rm log}~n )$ and at best near-linear in $n$ (i.e. $\approx \mathcal{O}(N_{\rm ref}~n)$).  In this work, our complexity is $\mathcal{O}(N_{\rm ref}~n^3 )$ per test event~\cite{Bonneel2011}.

This computational time cost can be converted into a memory cost by using Linearized OT (LOT) distances~\cite{Cai:2020vzx}. LOT takes advantage of the pseudo-Riemannian structure of 2-Wasserstein distances to simplify the calculation of OT distances. This is done by first defining the 2-Wasserstein tangent plane at a reference event, $\mathcal{R}$, living on the space of possible events. An event $\mathcal{E}$ can then be projected onto this plane by solving the OT problem between $\mathcal{R}$ and $\mathcal{E}$ to obtain the transport plan, $\gamma_{\mathcal{R}\mathcal{E}}$. After the events are embedded, simpler $L_2$ distances can be calculated between events on this tangent plane which faithfully approximate the 2-Wasserstein distance between events.  
Practically, $\mathcal{R}$ is chosen to be an unphysical reference event which adequately spans the ground space (i.e. a grid of $n$ point masses in the $d$ dimensional space). The OT distances between $\mathcal{R}$ and each of the $N_{\rm ref}$ physical reference events are precomputed and the $n \times n$ transport plans are stored, $\gamma_{\mathcal{R}\mathcal{E}_1}, ..., \gamma_{\mathcal{R}\mathcal{E}_{{N}_{\rm ref}}}$. For a given test event, $\mathcal{T}$, you then calculate one OT distance between $\mathcal{T}$ and $\mathcal{R}$ and store this transport plan, $\gamma_{\mathcal{R}\mathcal{T}}$. The last step is to calculate the Euclidean distance between   $\gamma_{\mathcal{R}\mathcal{T}}$ and each of the $N_{\rm ref}$ transport plans, $\gamma_{\mathcal{R}\mathcal{E}_1}, ..., \gamma_{\mathcal{R}\mathcal{E}_{{N}_{\rm ref}}}$. Thus, the time to evaluate each test event decreases from $\mathcal{O}(N_{\rm ref}~n^3 )$ to $\mathcal{O}(n^3~+~ N_{\rm ref}~n)$. However the memory cost is increased since the transport plans must be stored ($\mathcal{O}(N_{\rm ref}~n^2)$) rather than the point-mass coordinates themselves ($\mathcal{O}(n~d)$). Note that since LOT requires the transport plans themselves, distance-only approximation strategies like sliced Wasserstein distances cannot be used.

Additionally, under the min-OT strategy, we must also find the smallest of the  $N_{\rm ref}$ OT distances for each test event; with standard algorithms this has linear time complexity. 
Whereas, under the OT+ML strategies, this extra time cost is replaced by the time cost of the model (kNN or SVM) making its prediction.

\section{\label{sec:conclusion} Conclusion}

Optimal transport (OT) has been shown to perform well on event classification tasks for collider physics data~\cite{Cai:2020vzx, Cai:2021hnn, Craig:2024rlv},
which motivated this study in which we performed the first implementation of OT on neutrino LArTPC data.  In this work, we have applied OT to $e/\pi^0$ separation, a challenging pattern recognition task. We developed pre-processing techniques crucial for OT to be effective and  tested 2D and 3D OT, both on their own and as inputs for interpretable ML methods, including kNN and SVM. MicroBooNE Open Datasets~\cite{bib:uBopendataincl,bib:uBopendatanue} were used, which incorporate a realistic simulation of the MicroBooNE detector providing a robust framework to test performance. All methods significantly outperform the traditional reconstruction algorithm Pandora, and in particular combining 3DOT with SVM post-processing is shown to have the best performance out of all methods considered, with accuracy ranging from $74.2-87.6 \%$ and AUC ranging from $82.2-94.0\%$ depending on the energy bin. The fact that OT significantly outperforms traditional reconstruction methods in this task can be explained by its ability to quantify features that distinguish single-shower ($e$) vs. two-shower ($\pi^0$) events by utilizing the calorimetric charge distribution in the event. This bypasses  high-level EM shower reconstruction algorithms and hence avoids many bottlenecks and inefficiencies which have, to date, limited the performance of traditional methods.\\

There are a variety of future directions motivated by this work. The success of OT methods applied to $e/\pi^0$ particle classification
suggests applying OT to other reconstruction problems in LArTPC experiments such as $e/\gamma$ separation. The computational complexity of standard OT distance calculations warrants exploring strategies such linearized optimal transport (LOT)~\cite{Cai:2020vzx} for better computational efficiency. Finally, it would be worthwhile to explore the prospects for integrating OT into the reconstruction pipeline of neutrino LArTPC experiments comprising the SBN program and DUNE using a variety of charge readout methods. While this work has shown promising performance using wire-based TPC images, comparable performance is expected for strip-based detectors and OT may be even more suited to pixel-based readout thanks to its intrinsic 3D readout.

\section*{Acknowledgements}
We acknowledge the MicroBooNE Collaboration for making publicly available the data sets \cite{bib:uBopendataincl, bib:uBopendatanue} employed in this work. These data sets consist of simulated neutrino interactions from the Booster Neutrino Beamline overlaid on top of cosmic data collected with the MicroBooNE detector [2017 JINST 12 P02017].
We would like to thank Giuseppe Cerati for helping interpret the MicroBooNE public datasets and Pedro Machado for suggesting 2D projections.
This material is based upon work supported by the U.S. Department of Energy, Office of Science, Office of High Energy Physics under Award Numbers DE-SC-0023730 and DE-SC-0011702. Parts of this work were performed at the Kavli Institute for Theoretical Physics, supported by the National Science Foundation under Grant No. NSF PHY-1748958.
JNH was supported by the National Science Foundation under Grant No. NSF PHY-1748958, the Gordon and Betty Moore Foundation through Grant No. GBMF7392, and by a UCSB Chancellor's postdoctoral fellowship.

\appendix
\section{Gromov-Wasserstein Limited Run}\label{app:GWresults}
Gromov-Wasserstein distances~\cite{Mémoli_2011} were initially developed to calculate OT distances between distributions over different underlying metric spaces. For example, the OT distance between a 2D and a 3D distribution. To achieve this it relies only on relative distance information within each metric space. Concretely, let $\mu = \sum_i a_i ~\delta_{x_i}$ and $\nu = \sum_j b_j ~\delta_{y_j}$ be two discrete probability distributions over the ground spaces $X$ and $Y$, respectively. Let $D_{i,k} := d_X(x_i, x_k)$ where $d_X(\cdot, \cdot)$ is the Euclidean distance between points $x_i, x_k$ in the ground space $X$. Similarly, we have $D_{j,l} := d_Y(y_j, y_l)$. Then the $p$-Gromov-Wasserstein distance is given by
\begin{equation}
    GW_p(\mu, \nu) = \left( 
\underset{\gamma \in \Pi(\mu, \nu )}{\rm min} 
\sum_{i,j,k,l} \left|D_{i,k} -  D_{j,l} \right|^p
\gamma_{i,j} \gamma_{k,l}
\right)^{\frac{1}{p}},
\end{equation}
where $\Pi(\mu, \nu )$ denotes the set of all distributions over the product space $X \times Y$ with marginals $\mu,\nu$ (i.e. the set of all transport plans). Since Gromov-Wasserstein distances only rely on relative information, $D_{i,k}, D_{j,l}$, they are manifestly invariant to isometric transformations (i.e. rotations and translations) of points in the ground spaces. 

This useful feature comes at a price. Unfortunately, direct computation of Gromov-Wasserstein distances is a non-convex quadratic optimization problem, making it NP hard~\cite{kravtsova2025nphardnessgromovwassersteindistance}. While more-efficient approximation strategies exist~\cite{Peyr2016GromovWassersteinAO, vayer2022slicedgromovwasserstein}, Gromov-Wasserstein distances are still far more computationally expensive than classic Wasserstein distances.
This can be prohibitive for discrete distributions with many point masses.

Our hypothesis was that Gromov-Wasserstein distances might be particularly useful in this setting because they would obfuscate the need to align our 3D events in pre-processing. This possibility, and the fact that these could be run in off-line analyses, motivated us to explore this as an alternative.
We performed a limited run to see if the performance of Gromov-Wasserstein distances would out-compete our 3D alignment of events. We used the POT~\cite{flamary_2021} implementation of the entropically regularized algorithm described in Ref.~\cite{Peyr2016GromovWassersteinAO}.

\begin{figure}
    \centering
    \includegraphics[width=\linewidth]{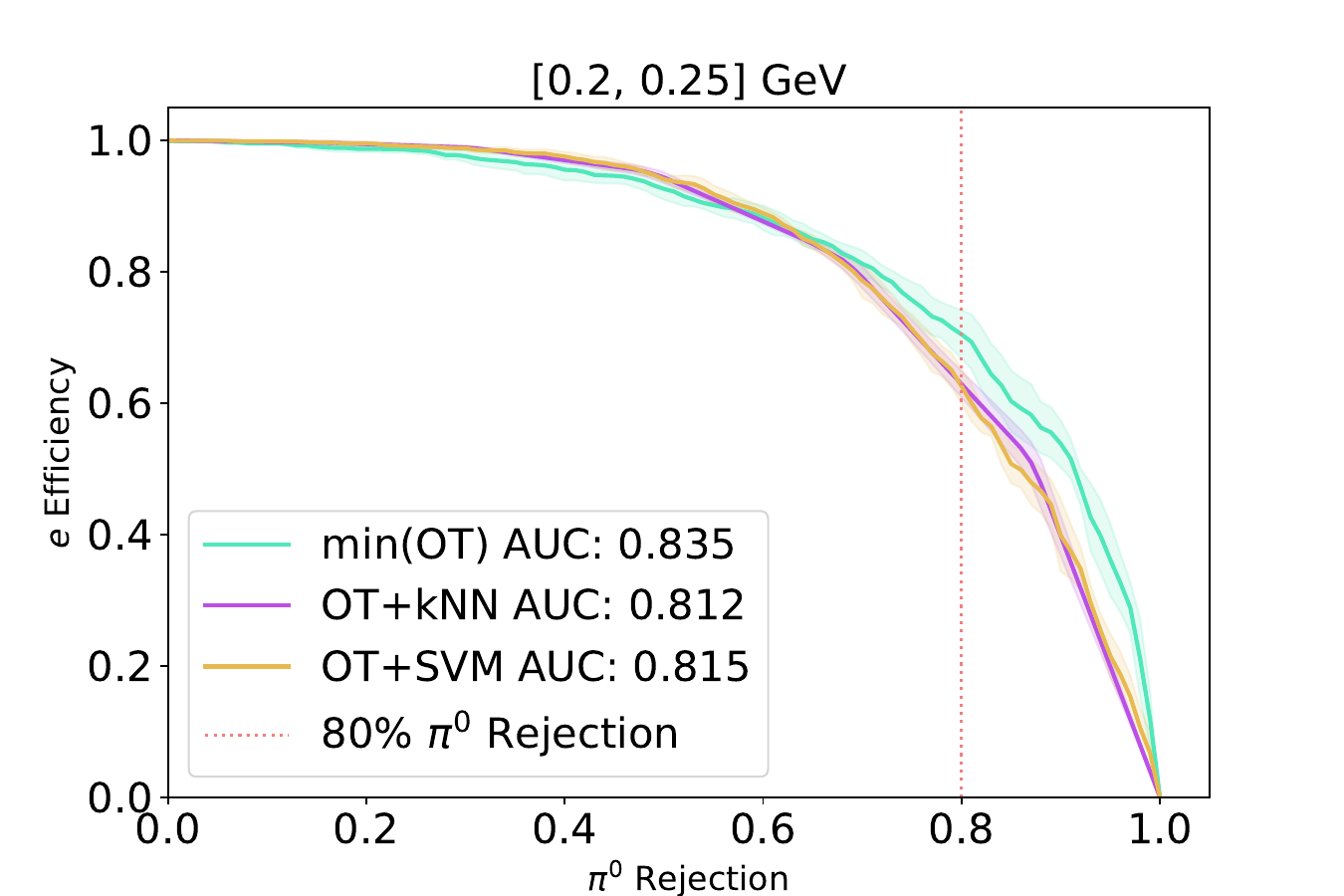}
    \includegraphics[width=\linewidth]{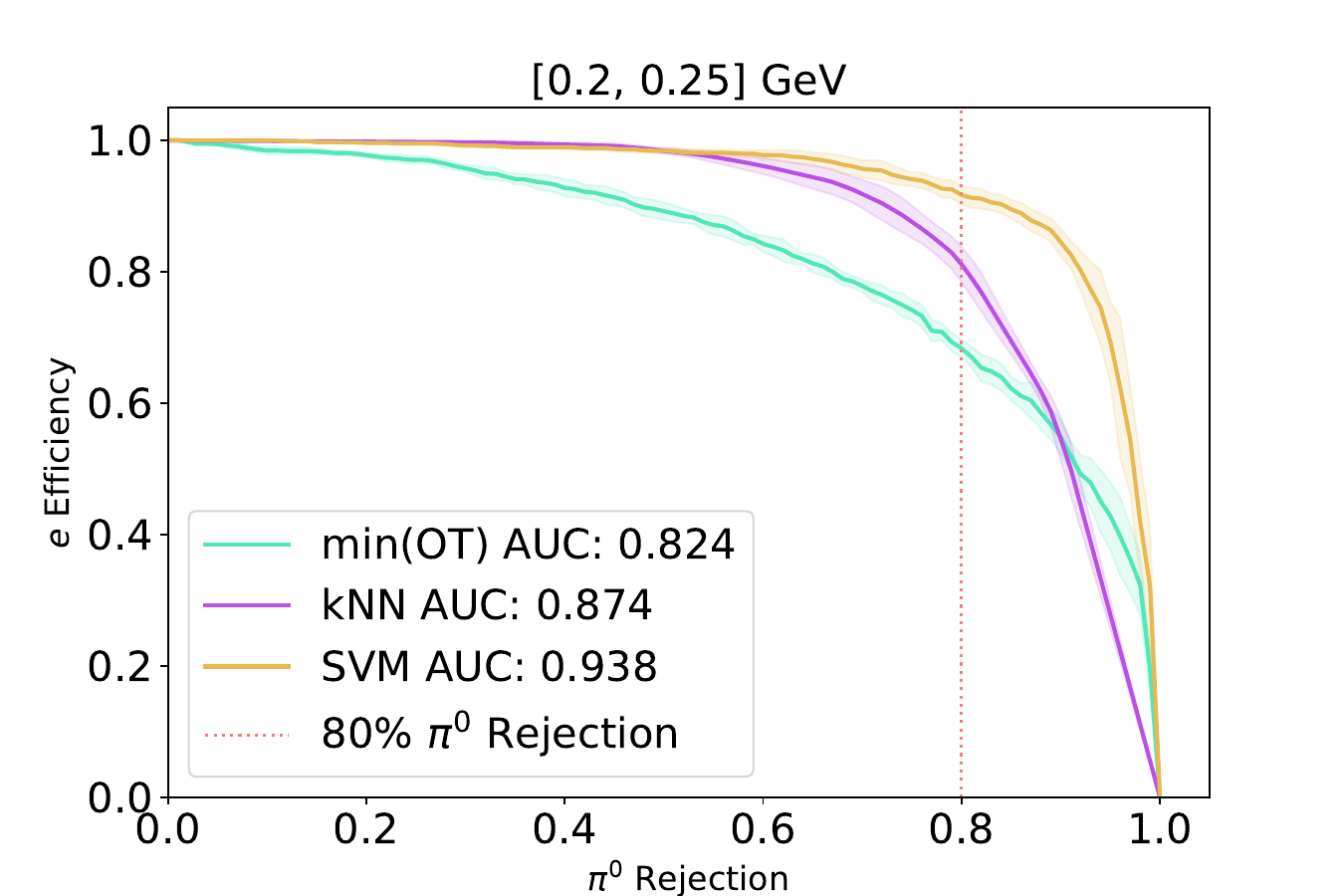}
    \caption{(Top) Results of a limited run to test the performance of Gromov-Wasserstein distances. (Bottom) The balanced 3D OT results on the same energy bin, $[0.2, 0.25]$, repeated from Fig.~\ref{fig:ROC} for comparison.}
    \label{fig:enter-label}
\end{figure}

We found that the min(OT) methods for both Gromov-Wasserstein and 3D balanced OT after alignment performed similarly, with Gromov-Wasserstein having a slightly higher AUC. This similarity gives us confidence that our pre-processing is accurately aligning these 3D events. Interestingly, when post-processing the Gromov-Wasserstein distances with interpretable machine learning methods (kNN and SVM) we do not see a significant performance jump. Whereas, in the balanced 3D OT case, we see a significant increase in performance. This could be explained by the fact that Gromov-Wasserstein solvers, like the ones used here, are approximate which may lead to in-exact representations of relative distances.

\section{Isolated Energy Deposit Removal}\label{app:AnomalyFiltering}
Certain events contain isolated small energy deposits which are far away from the main energy loss of the EM shower. These correspond to small branches of the shower produced by photons which travel a large distance before depositing their energy. Similar to the last section, the inclusion of these energy deposits may result in OT spending significantly more cost in the transport plan, leading to a high OT distance even between events of the same type.\\
To mitigate this effect, small isolated energy deposits are removed. A number of configurations are tested: clusters of isolated energy deposits are constructed by collecting together spacepoints within a range of distances (5, 8, 10, 12, 15, or 30 cm), and after this step those clusters with less than 3, 4, or 5 spacepoints are removed. No such configuration however leads to meaningful improvements in performance, leading us to avoid the isolated energy deposit removal for the main results presented here.

\bibliography{neutrinoOT}

\end{document}